\documentclass[twocolumn]{aa}

\usepackage{graphicx}
\usepackage{txfonts}
\usepackage{hyperref}
 \usepackage[normalem]{ulem}
 \useunder{\uline}{\ul}{}
 \usepackage{lscape}
 \usepackage{xcolor}

\begin{document} 


%

   \title{Multiple injections of energetic electrons associated with the flare/CME event on 9 October 2021}
   \subtitle{}

   \author{Immanuel. C. Jebaraj\inst{1,2,3}
     \and
          A. Kouloumvakos\inst{4}
     \and
          N. Dresing \inst{3}
     \and
          A. Warmuth \inst{5}
     \and
          N. Wijsen \inst{6,7}
     \and
          C. Palmroos \inst{3}
     \and
          J. Gieseler \inst{3}
     \and
          R. Vainio \inst{3}
     \and
          V. Krupar \inst{7, 8}
     \and
          J. Magdalenic \inst{1, 2}
     \and
          T. Wiegelmann \inst{9}
     \and
          F. Schuller \inst{5}
     \and
          A. F. Battaglia \inst{10,11}
     \and
          A. Fedeli \inst{3}
          }

\institute{Center for mathematical Plasma Astrophysics-CmPA, Department of Mathematics, KU Leuven, Celestijnenlaan 200B, B-3001 Leuven, Belgium.
        \and 
            Solar–Terrestrial Centre of Excellence—SIDC, Royal Observatory of Belgium, 1180 Brussels, Belgium
        \and
            Department of Physics and Astronomy, University of Turku, Finland.
        \and
            The Johns Hopkins University Applied Physics Laboratory, 11101 Johns Hopkins Road, Laurel, MD 20723, USA.
        \and
            Leibniz-Institut f\"ur Astrophysik Potsdam (AIP), An der Sternwarte 16, 14482 Potsdam, Germany. 
        \and 
            Department of Astronomy, University of Maryland, College Park, MD 20742, USA.
        \and
            Heliospheric Physics Laboratory, Heliophysics Division, NASA Goddard Space Flight Center, Greenbelt, MD 20771, USA.
        \and
            Goddard Planetary Heliophysics Institute, University of Maryland, Baltimore County, Baltimore, MD 21250, USA.
        \and
           Max-Planck-Institute for Solar System Research,
           Justus-von-Liebig-Weg 3,
           37077 G\"ottingen, Germany.
        \and
            Institute for Data Science (I4DS), University of Applied Sciences and Arts Northwestern Switzerland, Bahnhofstrasse 6, 5210 Windisch, Switzerland.
        \and
            Institute for Particle Physics and Astrophysics (IPA), Swiss Federal Institute of Technology in Zurich (ETHZ), Wolfgang-Pauli-Strasse 27, 8039 Zurich, Switzerland.
        }
            

\date{}

\titlerunning{Solar energetic electrons on 9 October 2021}

\abstract{We study the solar energetic particle (SEP) event observed on 9 October 2021, by multiple spacecraft including Solar Orbiter (SolO). The event was associated with an M1.6 flare, a coronal mass ejection (CME) and a shock wave. During the event, high-energy protons and electrons were recorded by multiple instruments located within a narrow longitudinal cone.}
{An interesting aspect of the event was the multi-stage particle energization during the flare impulsive phase and also what appears to be a separate phase of electron acceleration detected at SolO after the flare maximum. We aim to investigate and identify the multiple sources of energetic electron acceleration.}
{We utilize SEP electron observations from the Energetic Particle Detector (EPD) and hard X-ray (HXR) observations from the Spectrometer/Telescope for Imaging X-rays (STIX) on-board SolO, in combination with radio observations at a broad frequency range. We focus on establishing an association between the energetic electrons and the different HXR and radio emissions associated with the multiple acceleration episodes.}
{We have found that the flare was able to accelerate electrons for at least 20 minutes during the nonthermal phase observed in the form of five discrete HXR pulses. We also show evidence that the shock wave has contributed to the electron acceleration during and after the impulsive flare phase. The detailed analysis of EPD electron data shows that there was a time difference in the release of low- and high-energy electrons, with the high-energy release delayed. Also, the observed electron anisotropy characteristics suggest different connectivity during the two phases of acceleration.}
{} 


   \maketitle

%

\section{Introduction}

Acceleration of solar energetic particles (SEPs) during eruptive events may be associated with several different physical phenomena and mechanisms, such as solar jets and flares, CMEs, or shock waves \citep[e.g. see][]{Klein2017, Anastasiadis2019, Vlahos2019, Reames2021}, and they can accelerate particles to energies ranging from a few tens of keV to several GeV. The origins of SEPs measured in-situ have been a long-standing debate. This is because it is difficult to distinguish between several possible processes and interpretations using in-situ observations only near 1~AU. Previous studies suggest that both flare- and shock-related physical processes can contribute to the acceleration of SEPs \citep[e.g.,][]{Kouloumvakos2015,Papaioannou16}; however, it remains an open issue to quantify the contribution of each process to each species and a broad energy range. Energetic electrons are thought to be primarily accelerated in the low corona, and the physical mechanisms responsible for the acceleration of electrons may then be constrained to reconnection, for example at solar jets \citep[e.g. see][]{Krucker2011, Glesener2012, Musset2020, Zhang2022} or at current sheets forming at the wake of CMEs \citep{Kahler1992, Klein1999}. It is not yet well understood if shock waves can have any role in the acceleration at high energies (i.e. $>$1~MeV) through a drift acceleration at shock fronts \citep[][]{Ball2001}. 

While it is commonly believed that both protons and electrons can be accelerated to very high energies at magnetic reconnection sites in the low corona, it is still not clear which conditions can facilitate the escape of the accelerated particles from the acceleration site to the interplanetary space. The direct injection and escape of the energised electrons can be described in three scenarios, namely, the electrons released 1) by propagating shock fronts \citep[e.g.,][]{Kouloumvakos2022A&A}, 2) CME interaction with the ambient magnetic field lines, 3) by open magnetic field lines rooted directly to the active region that gives direct access to interplanetary space \citep[][]{Masson2019}. Depending on the proximity of the acceleration site to open magnetic field lines, the electrons can escape directly into the interplanetary space via open magnetic field lines and usually manifest as type III radio bursts \citep[see,][for a review]{Reid14Review}. 

Type III radio emission is produced when beams of energetic electrons resonantly generate Langmuir/slow-electrostatic waves which are linearly \citep[][]{Krasnoselskikh19, Tkachenko21, Jebaraj22} or non-linearly \citep[][]{Ginzburg58, Cairns87, Melrose17} transformed into electromagnetic waves at the plasma frequency and/or its harmonics. Type III radio bursts are known to be associated with flares, jets, and other solar eruptive phenomena. There is also a very good connection of energetic electron events with type III bursts at energies below 300~keV \citep[e.g.,][]{Krucker07,Krucker08,Klassen11a, Klassen11b, Klassen12, Klassen18}.

\begin{figure*}[ht]
\centering
\includegraphics[width= 0.99\textwidth]{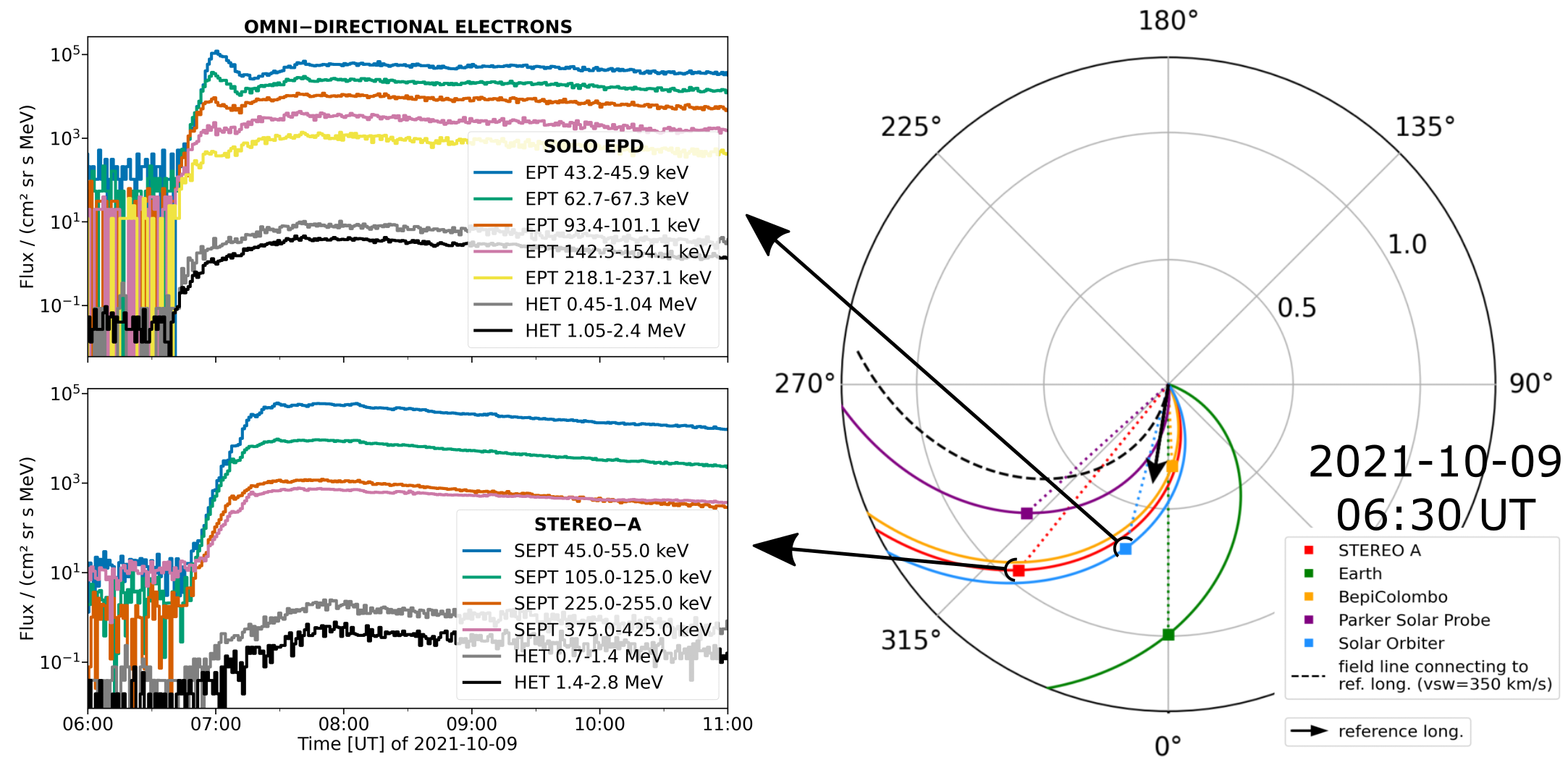}
\caption{The right panel shows a view of the heliographic equatorial plane from the north, in the Stonyhurst coordinate system, and the spacecraft constellation on 9 October 2021 at 06:30~UT. The different coloured squares indicate the location of all observing spacecraft, namely, L1 (Earth, green), \textit{STEREO-A} (red), \textit{SolO} (blue), and \textit{PSP} (purple). The nominal Parker spiral (curved line) and line of sight (dotted lines) from the Sun to each spacecraft is shown for each observer. The small back arrow and the dashed black line indicate the flare location and the reference Parker spiral. The large black arrows help locate the in-situ omni-directional electron recordings at \textit{SolO} and \textit{STEREO-A}, which are presented in the two rectangular panels on the left (note that \textit{STEREO-A}/HET data is not omni-directional, but pointing almost perpendicular to the nominal Parker spiral direction in the ecliptic plane; cf. Sec.~\ref{subsec:electron anisotropy}). \label{fig:Multy_sc_Electron} }
\end{figure*}

On the other hand, efficient shock acceleration of electrons, while possible, may be constrained to the shock drift acceleration mechanism \citep[SDA;][and references therein]{Ball2001} in regions where a quasi-perpendicular shock geometry is present. However, the efficiency of energy gain from the mechanism itself is rather limited since most electrons are transmitted through the shock transition regardless of electron energy when the shock geometry tends towards perpendicular to the upstream magnetic field ($\theta_{Bn} \approx $ 90$^{\circ}$). Evidence of shock waves accelerating beams of energetic electrons is common during solar eruptions and can be seen manifesting as drifting type II radio emissions. Numerous studies of type II radio bursts have shown that the emission most likely arises from the upstream regions of the shock wave, implying an active acceleration process at the shock wave \citep[][]{Krasnoselskikh85, Thejappa87, Jebaraj21, Kouloumvakos21} at multiple acceleration sites in some events. \cite{Aurass98} have suggested that coronal shock waves are able to accelerate electrons to energies considerably higher than the background thermal population.

A recent statistical study \citep[][]{Dresing22} has suggested that the acceleration of mildly-relativistic and relativistic electrons correlate rather well with critical shock parameters close to the Sun, namely, the shock strength (fast-magnetosonic Mach number). Similarly, good correlations have been found for high-energy protons as well \citep[][]{Kouloumvakos19}. This good correlation for electrons also suggests that efficient acceleration at shock regions with oblique and quasi-parallel shock geometry may also be present. However, the mechanism of electron acceleration in oblique shocks follows a diffusive shock acceleration mechanism \citep[DSA;][and references therein]{Bell78}, which is highly dependent on the shock wave's ability to accelerate ions \citep[for review; ][]{Treumann08para} and the subsequent generation of upstream wave turbulence. The electrons may then be trapped by the upstream waves and be accelerated through a Fermi acceleration mechanism similar to that of the protons \citep[e.g.,][]{Tsytovich73, Vaisberg83, Galeev84, Galeev95, McClements97, Gieseler00}.

In this study, we investigate an SEP event observed on 9 October 2021, by Solar Orbiter \citep[\textit{SolO;}][]{Mueller20}. The event was associated with a CME/shock wave and an M-class flare. During the event, high-energy protons and electrons were observed by multiple observers, such as, near-Earth spacecraft and Parker Solar Probe \citep[\textit{PSP;}][]{Fox2016}. Our motivation is to analyse what appears to be a second phase of electron acceleration as suggested by \textit{SolO} SEP electron observations, in combination with the complex radio observations that show multiple stages of particle energization during and after the flare impulsive phase. We find that this is an interesting aspect of the event, hence, our analysis is focused on establishing an association between the energetic electrons and the different radio emissions during the impulsive phase of the flare  and understanding the origin of the apparent second phase of electron energization after the flare maximum. Specifics on the energetic proton observations for this event can be found in \cite{Lario22}.

This study is organised as follows. We start with a brief introduction to the space-based and ground-based instrumentation and an overview of the event in Section 2 and 3, respectively. In Section 4, we show observations of the solar event with a special emphasis on the X-rays (Section 4.1), the radio waves (Section 4.2), and the solar energetic electrons (Section 4.3). In this section we also analyse the different observations. The results of our analysis and our conclusions about the two apparent phases of electron energization are provided in Section 5 and 6, respectively.

\section{Instrumentation}\label{Instrumentation}

For this study, we utilize data from instruments on-board \textit{SolO}, \textit{PSP}, Solar TErrestrial RElations Observatory Ahead \citep[\textit{STEREO-A}][]{Kaiser05, Kaiser08}, SOlar and Heliospheric Observatory \citep[\textit{SOHO};][]{Domingo95}, Solar Dynamics Observatory \citep[\textit{SDO};][]{Pesnell12}, and the \textit{Wind} \citep{Harten1995} spacecraft, as well as measurements from ground-based instruments. Below we give a summary of the data used in this study.

\begin{itemize}

    \item[\textbullet]
     \textit{Energetic Particle observations:}
     From \textit{SolO}, we utilize energetic particle measurements from different sensors of the Energetic Particle Detector \citep[EPD;][]{Pacheco2020} instrument suite, namely Electron Proton Telescope (EPT), High Energy Telescope (HET), and the Suprathermal Electrons and Protons (STEP), in the energy range from a few keV to a few MeV for electrons. From \textit{STEREO-A}, we utilize SEP measurements from the High Energy Telescope \citep[HET;][]{Rosenvinge2008} and the Solar Electron and Proton Telescope \citep[SEPT;][]{Mueller2008}. In addition, electron measurements from the 3DP \citep[][]{Lin1995} instrument on-board \textit{Wind} have been used.

    \item[\textbullet]
     \textit{Hard and soft X-rays observations:}
     We utilize hard X-ray (HXR) spectra and images from the Spectrometer/Telescope for Imaging X-rays \citep[STIX;][]{Krucker20} on-board \textit{SolO} and soft X-ray (SXR) observations from the Geostationary Operational Environmental Satellite \citep[\textit{GOES};][]{Howard94}.
    
    \item[\textbullet]
    \textit{Radio observations:}
    We utilize radio observations from both, ground-based and space-borne instruments. For the interplanetary part of the dynamic radio spectrum we utilize observations from the Radio and Plasma Waves instrument \citep[RPW;][]{Maksimovic20b} on-board \textit{SolO} and from the Radio Frequency Spectrometer \citep[RFS;][]{Pulupa17} part of the FIELDS electric antennas \citep[][]{Bale16} on-board \textit{PSP}. Observations from legacy instrumentation such as the SWAVES \citep[][]{Bougeret08} instrument on-board \textit{STEREO-A} and the WAVES experiment on-board the \textit{Wind} spacecraft \citep[][]{Bougeret95} are also employed. We also utilize ground-based radio observations from the Yamagawa radio spectrograph \citep[9 GHz - 70 MHz, ][]{Iwai17} and the e-Callisto network of radio telescopes \citep[in particular, ASSA (Astronomical Society of South Australia; 80 MHz - 16 MHz), ][]{Benz09}. These observations cover a broad range of the radio wavelengths, from the millimetric to the decametric domain.

    \item[\textbullet]
    \textit{Extreme ultra violet observations (EUV):}
     We utilize observations from the Atmospheric Imaging Assembly \citep[AIA;][]{Lemen12} on-board \textit{SDO} and the Extreme Ultra Violet Imagers (EUVI), part of the Sun Earth Connection Coronal and Heliospheric Investigation \citep[SECCHI;][]{Howard08} instrument suite on-board \textit{STEREO-A}.

    \item[\textbullet]
    \textit{Coronagraphs white light observations (WL):}
     We utilize coronagraphic observations in WL, from the Large Angle and Spectroscopic Coronagraphs (C2 \& C3) \citep[LASCO;][]{Brueckner95} on-board SOHO and the two coronagraphs (COR1 \& COR2), part of the SECCHI instrument suite on-board \textit{STEREO-A}.

\end{itemize}

\section{Event overview}

The event on 9 October 2021, was one of the first major eruptions of Solar Cycle 25 and was associated with an M1.6 flare in \textit{GOES} classification, a filament eruption, and a halo CME originating from the NOAA AR 12882. The active region had a $\beta \gamma$ configuration of its photospheric magnetic field \footnote{based on the classification provided in: \url{https://www.solarmonitor.org/index.php?date=20211009&region=12882}} and was located at the central meridian (N20W01). Soft X-ray (SXR) observations from \textit{GOES} showed that the solar flare started at 06:19~UT, peaked at 06:38~UT, and decayed to flux levels between the maximum and the pre-flare background level at 06:53~UT. The flare was also observed in hard X-rays (HXR) by the STIX instrument on-board \textit{SolO}, which at that time was at a heliocentric distance of 0.68~AU and was separated from the Sun-Earth line by 15.2$^{\circ}$ to the east (see Figure~\ref{fig:Multy_sc_Electron}). Hence, the flare was close to the disk centre for both \textit{SolO} and near-Earth assets. Radio observations were also very rich during the event, showing several complex spectral features. We provide details about the X-ray and radio observations in Sections~\ref{subSec:Xray} and \ref{subSec:Radio}, respectively.



\begin{figure}
\centering
\includegraphics[width= 0.49\textwidth]{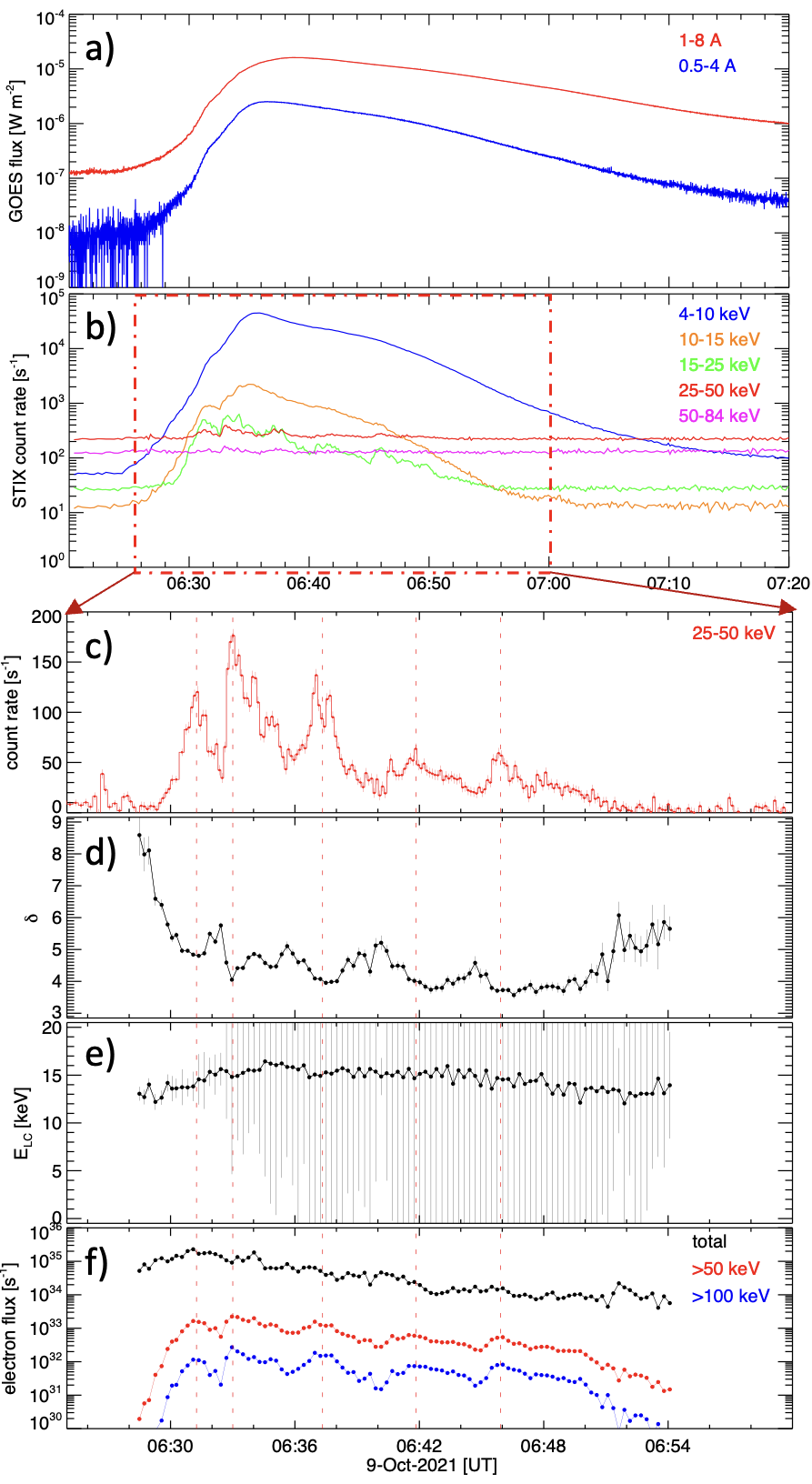}
\caption{(a) GOES soft X-ray fluxes shows the thermal emission of the M1.6 flare. (b) STIX hard X-ray count rates in five broad energy bands. The STIX times have been shifted to be consistent with the GOES observations from 1~AU. Results of the spectral fitting of STIX spectra show: (c) the background-subtracted STIX count rate in the 25--50~keV range, (d) the spectral index of the injected electrons, $\delta$, (e) the low-energy cutoff, $E_\mathrm{LC}$, and (f) the total injected electron flux, as well as the fluxes above 50 and 100~keV. The five major nonthermal peaks are indicated by red dashed lines. \label{fig:stix_goes_spectral} }
\end{figure}

\begin{figure*}[ttt]
\centering
\includegraphics[width= 0.99\textwidth]{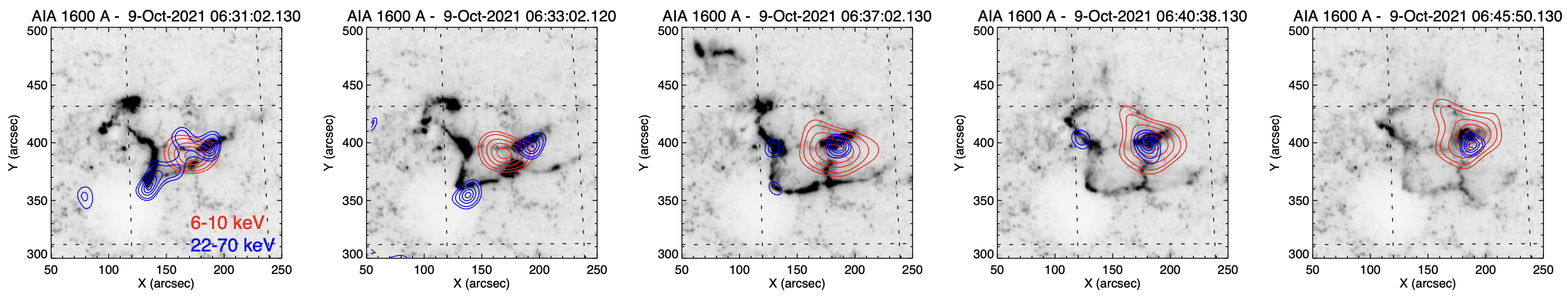}
\caption{Flare evolution as seen in a series of \textit{SDO}/AIA 1600~\AA\ images. Due to the inverted colour table, the flare ribbons and kernels are dark. The AIA frames have been rotated to conform to the view from SolO. The provided times represent the recording times of the AIA frames (UT at Earth). STIX images showing the thermal and nonthermal X-ray sources are overlaid as red and blue contours, respectively. \label{fig:stix_images} }
\end{figure*}

Figure~\ref{fig:Multy_sc_Electron} shows an overview of the energetic electron observations from \textit{SolO} (EPD-EPT) and \textit{STEREO-A} (SEPT) that observed the SEP event on 9 October 2021. All the spacecraft, and in particular \textit{SolO} and \textit{STEREO-A}, clearly observed an intensity increase of electrons at energies $>$1 MeV. The heliospheric view on the right panel of Figure~\ref{fig:Multy_sc_Electron} shows the heliographic equatorial plane from solar north and illustrates the spacecraft locations close to the start of the solar event, on 9 October 2021 at 06:30~UT. The new solar mission spacecraft (\textit{SolO} and \textit{PSP}), \textit{STEREO-A}, and near-Earth spacecraft were closely positioned in a quadrant and covered a narrow range of heliolongitudes of about 50$^\circ$. During the SEP event, \textit{SolO} (located at 0.68~AU), and \textit{STEREO-A} (located at 1~AU) were both trailing Earth for about 15$^\circ$ and $\sim$40$^\circ$, respectively. In the heliospheric view of Figure~\ref{fig:Multy_sc_Electron}, we also show the Parker spiral field lines connecting each spacecraft with the Sun. For the illustration of the spacecraft location, we used the Solar-MACH tool\footnote{https://solar-mach.github.io} \citep{Gieseler2022}. For each spacecraft, we used a solar wind speed of about 350~km/s to calculate the Parker spiral. We see that theinterplanetary magnetic field (IMF) lines connecting \textit{STEREO-A} and \textit{SolO} with the Sun are spatially close. We estimated the longitudinal difference of the magnetic footpoints at the solar surface to be around 4.9$^\circ$. From the electron recordings (left panel of Figure~\ref{fig:Multy_sc_Electron}, we see that at \textit{SolO} there is a rapid rise of the electron flux that lasts for about 15 minutes until the maximum, and then there is a second more gradual increase after the maximum. This second increase, however, was not observed by the closely connected \textit{STEREO-A} spacecraft.

\section{Observations and data analysis}\label{Observations}

\subsection{X-ray observations} \label{subSec:Xray}

Figure~\ref{fig:stix_goes_spectral} (panels (a) and (b)) shows the GOES soft X-ray fluxes in comparison with the STIX hard X-ray count rates in five wide energy bands (note that 160~s have been added to the STIX times to account for the light travel time difference between \textit{SolO} and \textit{GOES}). While the count rates below 15~keV show the smooth time evolution characteristics for the thermal bremsstrahlung produced by hot plasma, the multiple spikes visible at higher energies suggest several episodes of electron acceleration. To investigate this quantitatively, we have forward-fitted a time series of observed STIX count spectra with the combination of an isothermal model and a thick-target nonthermal component \citep{Brown71}, using the OSPEX spectral analysis tool\footnote{http://hesperia.gsfc.nasa.gov/ssw/packages/spex/doc/}.

The fit results for the nonthermal component are shown in Fig.~\ref{fig:stix_goes_spectral} (panels (c), (d), (e), and (f)). Similarly to panels (a) and (b) of the same figure, the times have been shifted to be consistent with a viewpoint at 1~AU. The background-subtracted count rates in the range of 25--50~keV  (panel (c)), show at least five major peaks, indicated by dashed vertical lines. The spectral index $\delta$ of the injected electron flux, in panel (d), shows a clear anti-correlation with the nonthermal count rates. This is the well-known soft-hard-soft behaviour \citep[e.g.,][]{Grigis04}, and it is indeed present in each peak. Overall, the spectrum gets harder over time (soft-hard-harder evolution). This behaviour was observed in some solar flares \citep{Grigis08}, particularly in those associated with interplanetary proton events \citep{Kiplinger95}. Next, panel (e) shows the low-energy cutoff $E_\mathrm{LC}$, which lies mostly in the range of 12--16~keV. This quantity is rather ardous to constrain (note the large uncertainties), and we can only determine the highest $E_\mathrm{LC}$ that is consistent with the data because the true cutoff is usually masked by the thermal emission \citep[see e.~g.][]{Warmuth20a}. Adopting this, panel (f) shows the total injected electron flux, which reflects a lower estimate for this quantity. Additionally, we show the fluxes above 50~keV and 100~keV in red and blue, respectively. Note that the HXR count rate peaks are indeed associated with higher electron fluxes, which are most clearly seen at higher energies.

To investigate electron acceleration in more detail, we performed image reconstruction of the hard X-ray sources using the STIX pixelated science data \citep[cf.][]{Krucker20,Battaglia21,Massa22}. Figure~\ref{fig:stix_images} shows the evolution of the thermal (6--10 keV, red contours) and nonthermal X-ray sources (22--70 keV, blue contours) reconstructed with the MEM\_GE imaging algorithm \citep{Massa20}. Countour levels start at 20\% of the maximum intensity. The image times correspond to the five major nonthermal peaks. While the integration time was one minute for the thermal images, it varied between one and four minutes for the nonthermal peaks in order to optimize counting statistics. The sources are overplotted on AIA 1600~\AA\ images that have been rotated to conform with the STIX viewpoint \citep[see][]{Battaglia21}. The inverted colour table shows the flaring ribbons and kernels in dark. The UV images show that the flare had an unusually complex morphology with multiple ribbons and kernels, lying partly at right angles to each other.

The X-ray sources were co-aligned with the rotated AIA images using the pointing information provided by the STIX Aspect System \citep[SAS;][]{Warmuth20b}, which has a nominal positioning accuracy of $\pm$4". We see that during the first two nonthermal peaks, two HXR footpoints (FPs) are located at the southern end of the eastern flaring ribbon and the western flaring kernel. At the second peak, an additional weaker FP appears near the central part of the eastern ribbon. At the third nonthermal peak, the southeastern FP has faded away, and we have again the classical two-footpoint configuration, which is also maintained during the subsequent two peaks. While the western FP remains stationary, the eastern FP moves along the flare ribbon to the northeast. Note that the eastern FP has nearly faded away in the fifth peak, so that in contrast to the other panels, the nonthermal contour lines shown here start already at the 5\% level. 

In an effort to visualise the complex STIX observations with respect to the magnetic field topology of the active region, we have performed a non-linear force free field extrapolation. The details of the analysis are provided in Appendix. \ref{appendix:NLFFF}. The HXR sources and their movement correspond well with the western part of the AR and the dominant footpoint rooted in the positive polarity region. The shifting of the HXR footpoints is seen in the south-eastern parts of the AR. 

In summary, we conclude that the flare was able to accelerate electrons from at least 06:28 to 06:54~UT, mostly in the form of five discrete pulses. A change of magnetic topology appears to have taken place between the first and third nonthermal burst, shown by a significant shift of the nonthermal emission in the eastern part of the flare.


\subsection{Radio observations} \label{subSec:Radio}

\begin{figure*}
\centering
\includegraphics[width= 0.99\textwidth]{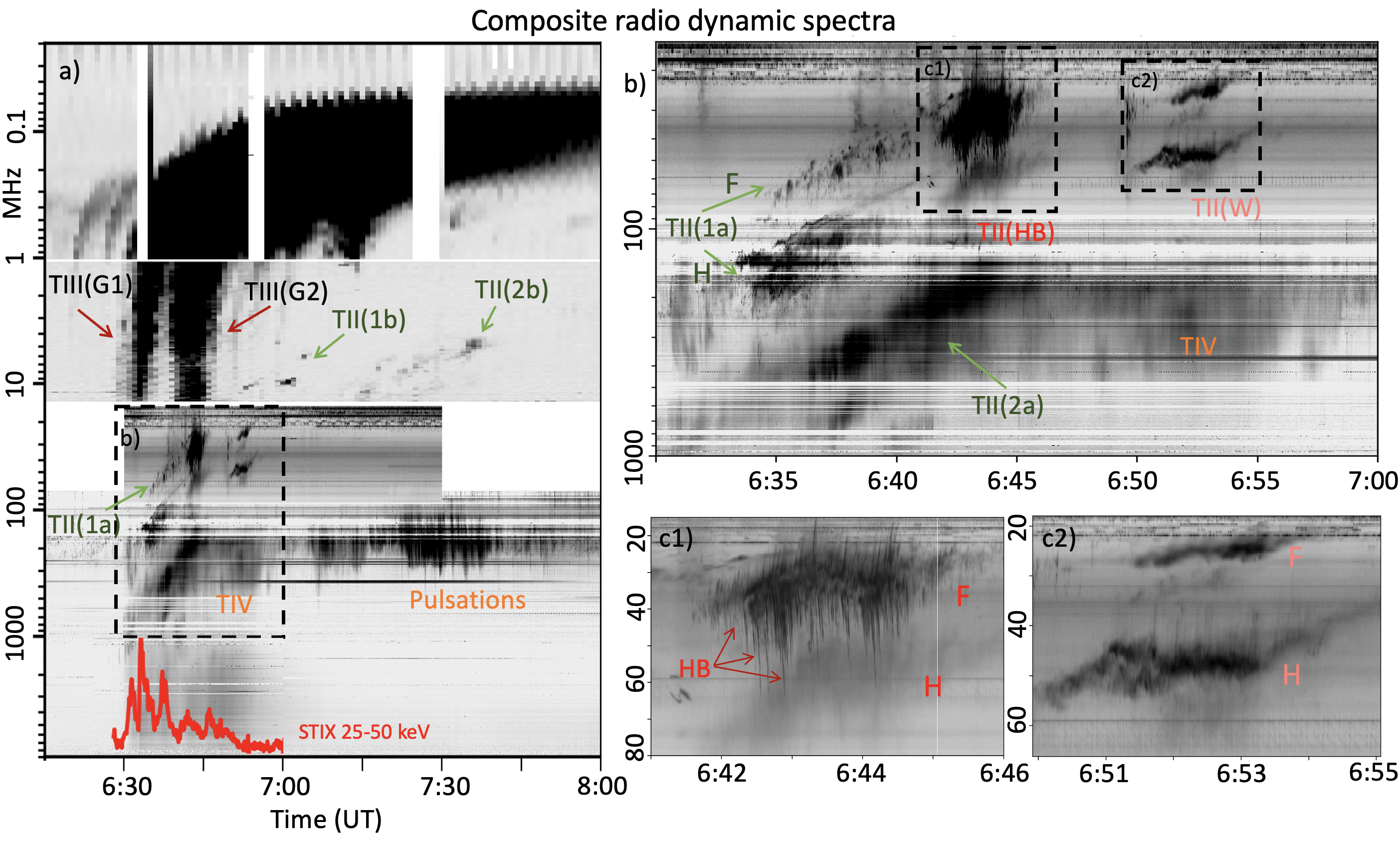}
\caption{The radio event associated with the flare/CME on 9 October 2021. Panel (a) shows the full radio spectrum from the Earth vantage point including both ground-based and space-borne observations in the range of 10 GHz to 20 kHz. The different types of radio bursts are marked on the spectra with their respective abbreviations. Panel (b) presents a zoomed in view of the decimetric-metric-decametric observations that showcase different structured radio emission, such as herringbones and a stationary flare continuum. The details of the structured radio emission, i.e. type II herringbones(TII(HB)) and very narrow-band wavy part of the type II (TII(W)), are presented in panels (c1) and (c2), respectively. \label{fig:DH_radio} }
\end{figure*}

In Figure~\ref{fig:DH_radio}, we show a composite dynamic radio spectrum constructed from the observations of several ground-based and space-borne instruments. Such a composite spectrum provides uninterrupted high time and frequency resolution observations, starting from microwave wavelengths that are generated at altitudes very close to the chromosphere and up to the kilometric wavelengths, correspond to the interplanetary space. The combined observations indicate a wide variety of coronal processes related to the acceleration and propagation of nonthermal electrons.

The solar radio event is rich with different types of radio emissions i.e., type II, III, radio bursts, IV continuum. In microwave wavelengths (9 GHz -- 1 GHz) we observe a diffuse radio emission which is most-likely gyro-resonant in nature \citep[e.g.,][]{Bastian98} and is emitted by near-relativistic electrons ($\approx$100 keV) trapped in the flaring loops \citep[e.g.,][]{Nindos20}. The low-decimetric to decametric wavelengths (i.e., from $\sim$1 GHz to $\sim$20 MHz) are mostly dominated by plasma emission (e.g., type II, III, IV radio emissions) produced by fast electron beams that are associated with flares, propagating shock waves, and electrons trapped within flare loops or in flux rope CMEs \citep[][]{Mclean85}. From the composite dynamic radio spectrum of these wavelengths (Figure~\ref{fig:DH_radio}), we distinguish different type III radio bursts, some of them associated with the HXR pulses, multiple parts of type II burst that exhibit various fine structures, and a type IV continuum. At hecto-kilometric wavelengths, the interplanetary radio emissions associated with the event were observed by all spacecraft namely, \textit{PSP}, \textit{SolO}, \textit{STEREO-A}, \textit{Wind}. We observe two main type III radio burst groups and two patchy parts of type II radio burst.

The event of 9 October 2021, was first observed in radio wavelengths with the emission starting at around 06:30~UT. The diffuse radio emission observed in microwave wavelengths lasted for the entire duration of the flare and corresponds well with the HXR pulses as shown in Fig.~\ref{fig:stix_goes_spectral}, from 06:30~to 06:54~UT. At the beginning of the event, we observe the first type III radio burst (TIII-G1), which was associated with the first peak in HXR during the flare impulsive phase. The first properly distinguishable signature (although faint) of G1 was observed at 06:31:40~UT, starting at $\sim$40~MHz. Apart from the TIII-G1 and the microwave continuum, near the very beginning of the event, we observe structured and narrowband spike-like emission at around 350~MHz, at 06:30~UT. Similar type of emission was also observed at higher frequencies, around 1000 MHz, at the same time. The morphology and the apparent drift of the features suggest that they could be precursors \citep[][]{Farnik03, Pohjolainen08b} to the type II radio burst observed a few minutes later.

The first part of the type II (TII(1a)) radio burst was observed at 300 MHz around 06:33~UT. This 2nd harmonic emission lane ($2f_{pe}$) shows rather patchy morphology. The fundamental lane ($f_{pe}$) of the same type II was observed later (06:34:20~UT) starting at about 90 MHz, and it was considerably patchier than the second harmonic and consisted of distinguishable fine structures. The nature of the narrow-band patchy features indicates rather localized source regions with rapidly changing plasma conditions. Such morphological characteristics are common in the case of metric-decametric type II bursts \citep[e.g.,][]{Cairns03, Kouloumvakos21, Jebaraj21}. The harmonic component was observed at $\sim$250 MHz meaning that a shock wave was formed at $\sim$1.4 $R_{\odot}$ \citep[applying a 2-fold Newkirk coronal electron density model; ][]{Newkirk61}. The coronal electron density is much higher in the corona close to the flaring region and at regions that are probably dominated by closed magnetic field lines. The radio emission ended at around 06:42~UT, and at 23 MHz and 50 MHz for the fundamental and harmonic lanes, respectively.

\begin{figure*}[t]
\centering
\includegraphics[width=0.99\textwidth]{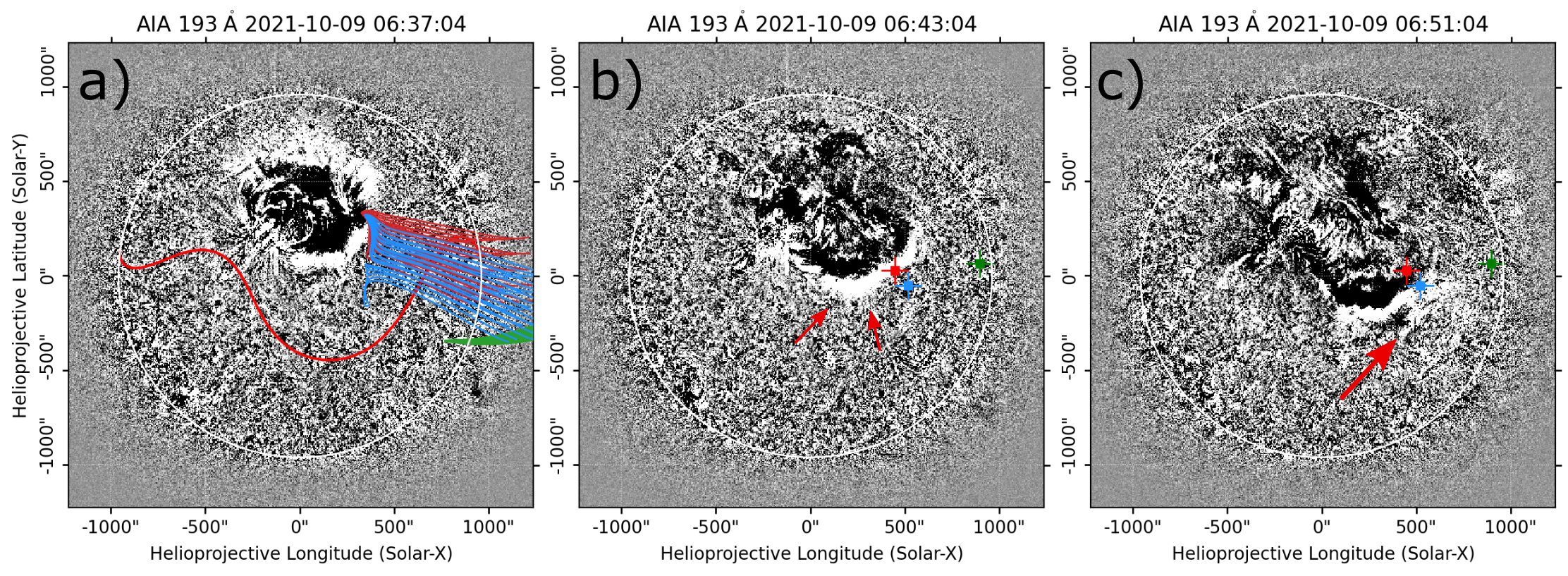}
\caption{The evolution of the EUV wave as observed by \textit{SDO}/AIA 193 $\AA$ at three time instances. Panel (a) shows the EUV wave close to the SXR peak time. The open magnetic field lines connecting to both \textit{SolO}, \textit{STEREO-A}, and \textit{Wind} are drawn in blue, red, and green, respectively. The red line across the Sun face represents the heliospheric current sheet. Panels (b) and (c) show the EUV wave expansion at 06:43~UT and 06:51~UT roughly corresponding to TII(HB) and TII(W). The red arrows indicate the region of interest where the wave also undergoes reflection at the coronal hole boundary. The blue, red, and green markings in (b) and (c) represent the spacecraft connections shown in (a). \label{fig:EUV_wave} }
\end{figure*}

Recently, \cite{Ramesh22} imaged the 80 MHz harmonic component of TII(1a) at $\approx$ 06:38~UT, which corresponds to the end of the inferred injection time of the low-energy electrons. The source of TII(1a) according to this study was located close to the southwest periphery of the flaring active region. This position roughly coincides with the direction of the strong EUV wave expansion, toward the southwest direction from the source region, where the open magnetic field lines were connected to \textit{SolO} and \textit{STEREO-A} (Sec.~\ref{fig:EUV_wave}).


Simultaneously with the TII(1a), we traced also a second, more diffuse and broadband radio emission possibly also type II burst (TII(2a)). This emission lane had a comparable drift rate to the one of TII(1a), and it was observed starting at high frequency $\sim$600 MHz, at around 06:36~UT. Approximately at the same time, also a continuum emission was observed, starting at $\sim$1 GHz (at around 06:31~UT). The continuum emission seems to be superposed with the TII(2a) burst. Such a synchronous appearance makes the separation of the two different types of radio emission very difficult. A continuation of the broad-band continuum was also observed after the cessation of TII(2a), but with variable intensity.  The intense part of the broadband continuum with fine structures (marked in Fig.~\ref{fig:DH_radio}a,b as TIV) started at 06:47~UT. The stationary TIV extended from $\sim$600 MHz to 200 MHz and consisted of intense broadband pulsation, extending till about 08:00~UT.

The second and most intense group of TIII-G2 radio burst was observed after the flare impulsive phase at around 06:37~UT, starting at $\sim$70 MHz and corresponding well to one of the HXR pulses. We find no clear evidence to confirm that the type III bursts emanate from the type II burst, so called type II-associated bursts \citep{Cane81,Dulk2000}. However, concurrently with the appearance of the TIII-G2 radio burst a clear brightening of a type II was observed. The distinguishable herringbone features (TII(HB)) at the decametric wavelengths (40 MHz fundamental emission) can be observed. TII(HB) was observed together with TIII-G2 starting from 06:41~UT and marks a period of very efficient electron acceleration at the shock wave. It is possible that some of the electron beams generating TIII-G2 were accelerated by the shock wave. We show the details of the corresponding dynamic spectrum in Figure~\ref{fig:DH_radio}.(c1). The fast drifting herringbone bursts originating from both sides of the type II backbone are generated by the fast electron beams accelerated in the upstream region of the shock wave. The geometry of the shock wave in these regions is close to $\theta_{Bn} \approx$ 90$^{\circ}$, which is evident by the lack of a bright backbone \citep[][]{Stewart80}. The drift rate of the herringbone features and the irregular morphology of the backbone may reflect the physical characteristics of either the shock wave or the variations in the local plasma frequency close to the shock wave. The spectral irregularities of the backbone structure are somewhat to be expected when the shock wave interacts with an electron-rich environment (e.g., a streamer) and therefore a reverse drift into higher plasma frequency.  

A few minutes ($\sim$5 min.) after the end of the TII(HB), we observed a similar (wavy) type II-like burst (TII(W)). The TII(W) burst starts at 06:50~UT, and as it is unclear if it is the decametric continuation of TII(2a) ,we distinguish it just as TII(W). A zoomed-in view of TII(W) is presented in Figure~\ref{fig:DH_radio}.(c2), which shows that the second harmonic of the emission was brighter than the fundamental, which is opposite to the fundamental brightening in the case of TII(HB). While TII(W) also appears to have some herringbone-like features, they are not as clear as those observed in TII(HB). Another important characteristic of TII(W) is also the presence of a bright backbone together with the herringbones, indicating a shock wave that is probably not as perpendicular as TII(HB).


The two main parts of the type II radio burst, namely TII(1a) and TII(2a), observed in the metric wavelengths were also observed in the hectometer wavelengths. The hectometric counterpart of the two parts of the type II burst were also noticeably patchier and bursty compared to their decametric counterparts. The continuation of the two parts of the metric type II radio burst was recorded by the space-based instruments and it is marked as TII(1b) and TII(2b) in Fig. \ref{fig:DH_radio}. 

The TII(2b) was observed in the hectometer range starting at 7:08~UT, which was about 12 minutes after the appearance of the TII(W). The relative bandwidth of the hectometric type II burst was comparable to that of the decametric, suggesting that the source of the type II bursts may have been located in a relatively small region of the shock wave \citep[][]{Schmidt2016, Kouloumvakos21, Jebaraj21}. Both bursts, namely, TII(1b) and TII(2b) come to a cessation at 07:07~UT (6~MHz) and 07:40~UT (4.5~MHz), respectively. 

Around the same time that the type II radio emission appeared in the hectometric wavelengths, we continued to observe also the TIV with sporadic broad-band radio pulsations in the metric wavelengths centred at 150 MHz. This emission was most likely produced by electrons trapped in post-flare loops.

Hecto-kilometric observations of the two type III groups, namely, TIII-G1 and TIII-G2 were recorded by all available space-borne observers. At kilometric wavelengths, both TIII-G1 and G2 were also observed together with Langmuir waves at \textit{SolO} and \textit{STEREO-A}. Langmuir waves are fundamental plasma waves that are generated by the electron beam, which can be subsequently converted into type III radio emission observed at large angles \citep[][]{Jebaraj22}. In-situ Langmuir waves are an indication of the electron beams propagating at close proximity to the observing spacecraft. Despite the fact that all the spacecraft observed the different type III bursts, only \textit{STEREO-A} and \textit{SolO} observe Langmuir waves. This confirms the passage of the type III generating electron beams through the position of the spacecraft and indicates that the magnetic connectivity of the two spacecraft with the electron beam was most likely similar. On the other hand, \textit{PSP} and \textit{Wind} did not observe Langmuir waves, and this indicates that the electron beams did not propagate in their vicinity. An in-depth analysis of the directivity of both TIII groups is presented in Appendix~\ref{Sec:appendix_radio}.


\subsection{Remote sensing observations in EUV and WL} \label{subSec:Shockwave}

The solar event on 9 October 2021,  was associated with a CME and a large-scale propagating coronal wave that was observed in both EUV and WL. While there has been an extended debate on the physical nature of such large-scale coronal disturbances \citep[for a review, see][]{Warmuth15}, they are now generally interpreted as signature of fast-mode waves or shocks \citep[e.g.][]{Long17}. The expansion of the EUV wave was first observed by \textit{SDO}/AIA at 06:30~UT. Figure~\ref{fig:EUV_wave} presents \textit{SDO}/AIA running difference images at 193~$\AA$, from different times of the EUV wave evolution. Due to the presence of a small coronal hole and a streamer south of the flaring active region, the evolution of the wave was rather constrained and deformed in the low corona. The presence of pre-existing magnetic structures such as coronal holes and streamers can affect the propagation and characteristics of large amplitude waves \citep[][]{Vrsnak00a, Vrsnak00b}. This mainly occurs because, in these regions, there is a rapid change in the characteristic speed of the medium (e.g., the fast-magnetosonic speed). The evolution of the low-coronal EUV wave shown in Fig.~\ref{fig:EUV_wave} was observed to be rather strong in the southwest quadrant away from the eruptive source region. Such an intense evolution of the EUV wave in this direction was also accompanied by a propagating pressure wave at higher altitudes, which was observed by the coronagraphs.

We utilise the information from radio observations in tandem with EUV images to understand the evolution of the pressure wave and the formation of a shock wave in the low corona. The presence of multiple type II radio bursts (described in Sec.~\ref{subSec:Radio}) suggests that a shock wave was present very early during the event at multiple locations. TII(1a) was observed at a lower frequency, while TII(2a) was observed at a much higher frequency 3 minutes later ($\approx$ 06:36~UT, Sec. \ref{subSec:Radio}). Such a disparity in starting frequencies suggests that the source of TII(1a) was possibly closer to the leading edge of the pressure wave, while the source region of TII(2a) could have been located closer to the flanks. Near the start of both TII(1a) and TII(2a), the propagation and interaction of the wave with closed field magnetic structures at the periphery of the active region most likely led to favourable conditions for electron acceleration and the subsequent generation of type II radio emission in regions close to the apex and the flanks of the wave \citep[e.g.,][]{Kong16,Kong17,Kouloumvakos21}. 

After 06:41~UT, the wave evolution was rather strong in the southwest direction (Fig.~\ref{fig:EUV_wave}a). Close to the start of the  TIII-G2 and also TII(HB), we observed an interaction of the EUV wave's southwest flank with open field lines that are probably part of a streamer. In Fig.~\ref{fig:EUV_wave}a, we show the open magnetic field lines that connect to \textit{STEREO-A}, \textit{SolO}, and Earth. These field lines were derived using the Potential Field Source Surface (PFSS) model \citep{Schrijver03} and Air Force Data Assimilative Photospheric Flux Transport (ADAPT) magnetograms \citep{Arge2009}. The interaction of the pressure wave with the magnetic structures in this region was most likely an important catalyst for efficient electron acceleration that we observed in radio wavelengths (TIII-G2 and TII(HB) in Fig.~\ref{fig:DH_radio}c1) and in-situ energetic electrons observed by \textit{SolO} and \textit{STEREO-A}. Furthermore, during the time of these interactions, both spacecraft were connected to the EUV wave.

When the TII(HB) first appears in the radio spectrum at 06:41~UT, we observe that the EUV wave propagated past the streamer towards the boundaries of a coronal hole southwest of the active region. This area is marked roughly by the red arrows in Fig.~\ref{fig:EUV_wave}b at 06:43~UT. When the wave interacted with this coronal hole, part of the wave was reflected by the coronal hole boundary at $\sim$06:50~UT (Fig.~\ref{fig:EUV_wave}c). A faint part of the wave transmitted across the coronal hole boundary and also refracted into regions where it was more favourable for the shock to exist, i.e., higher in the middle corona where the local speed of the medium decreases \citep[][]{Uchida73, Warmuth05}. Due to the low density of the coronal hole region and the high magnetic field strength (increased characteristic speed of the medium), the large amplitude wave may propagate faster and also dampen at an increased rate. During and well after this part of the interaction at the coronal hole boundary, we observed the TII(W) presented in Fig.~\ref{fig:DH_radio}.c2. 

In WL, the pressure wave was first observed by \textit{SOHO}/LASCO C2 coronagraph at 07:12~UT as a halo-shaped coronal wave propagating higher in the middle solar corona along the position angle (PA) 263$^{\circ}$ (i.e., in the south-west quadrant). This part of the WL shock wave corroborates well with the fastest component of the EUV wave as we discussed earlier. The observations also validate the aforementioned refraction of the pressure wave higher in the corona, where there may exist more favourable conditions for the formation of a shock wave (e.g., lower Alfven speed). As the event originated close to the central meridian, the pressure wave was observed as a halo event encircling the solar disk from the Earth's point of view. Due to projection effects, tracking the shock wave becomes increasingly difficult at later times in both LASCO C2 and C3 coronagraphs.

At \textit{STEREO-A}, the event was observed $\sim$40$^{\circ}$ closer to the west limb. The pressure wave was first observed in the COR1 coronagraph at 06:46:18~UT and in the COR2 coronagraph at 07:08:45~UT. At the time when the pressure wave enters the COR1 field of view, all the observed radio emission was in the decametric wavelengths (mid-coronal plasma) and the TII(1a), TII(HB), and TIII-G2 were near their cessation. This shows that the most efficient electron acceleration took place when the shock wave was still located low in the middle corona. The pressure wave evolved fastest along PA 266$^{\circ}$ which suggests that this was almost the same leading-edge identified in the LASCO C2 observations. The wave was highly deformed by the presence of two streamers at PA 230$^{\circ}$ and at PA 280$^{\circ}$ further suggesting the interaction between the pressure wave and the various density structures on the Sun. These interactions may also have contributed to the patchy and bursty yet continuous observations of the TII(1b) and TII(2b) at hectometric wavelengths until their cessation at 07:07~UT and 07:40~UT, respectively. The kinematics of the shock wave and their association with TII(1a), TII(ab), TII(2a), and TII(2b) are presented in Appendix~\ref{Sec:appendix_shock}. 


\subsection{Energetic electrons} \label{subSec:EEtiming}

\subsubsection{Pitch-angle distributions} \label{subsec:electron anisotropy}

\begin{figure*}[ht]
\centering
\includegraphics[width= 0.99\textwidth]{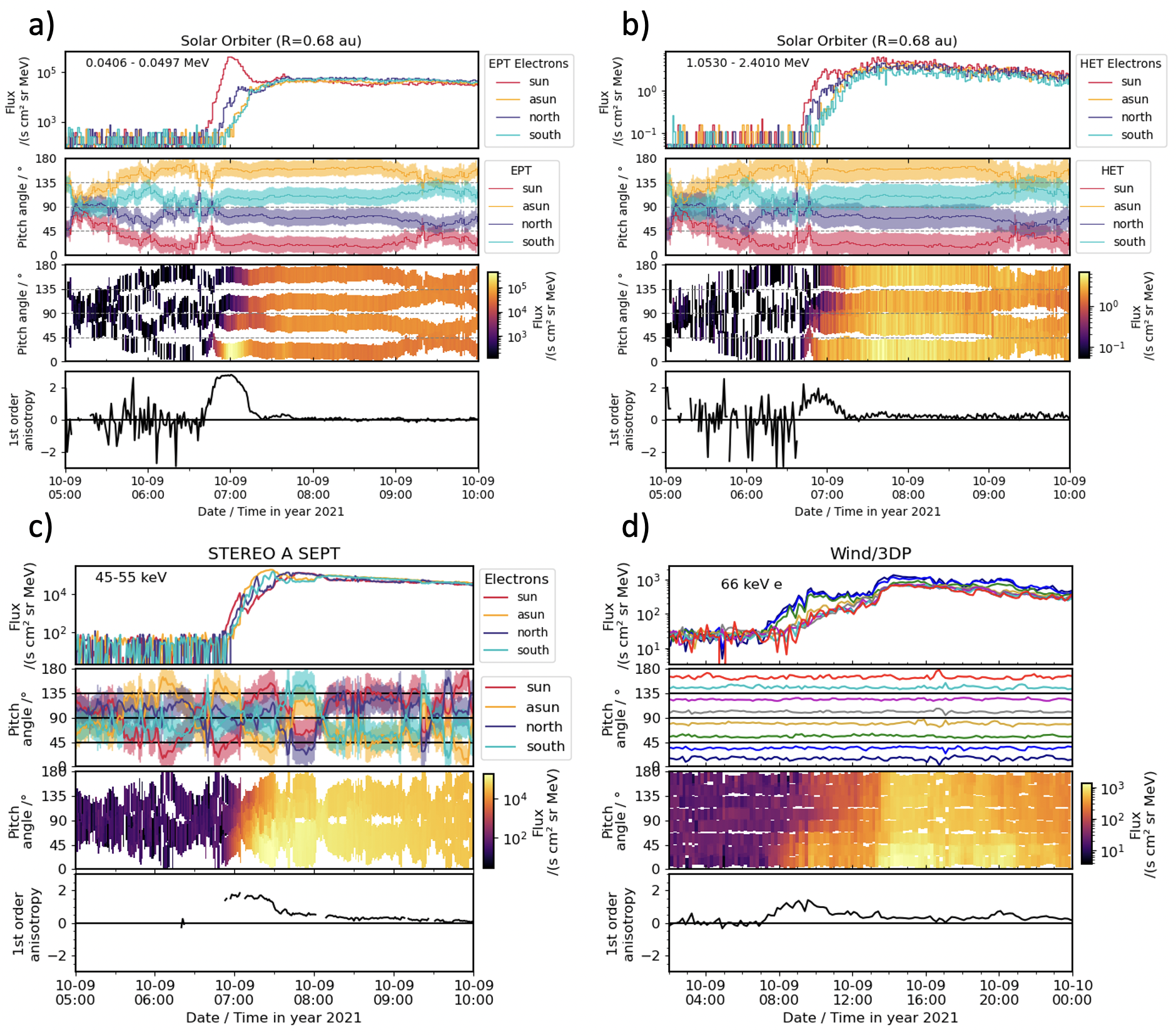}
\caption{In-situ electron recordings by multiple observers. Energetic electron event observed by \textit{SolO}/EPT (a) and HET (b). The top panels of (a) and (b) show the intensities of 40--50~keV and 1.1--2.4~MeV electrons as observed by the four different sensors of the EPT and HET instruments. The second panels show the corresponding pitch-angle coverage, and the third panels show the pitch-angle dependent intensities with intensity level marked by colour coding. The bottom panels show the first order anisotropies. Energetic electron event observed by \textit{STEREO-A}/SEPT in the 45--55~keV energy channel (c) and \textit{Wind}/3DP (d) in the 50--82~keV channel. The panels have the same content as for (a) and (b), but (d) shows a longer time period. \label{fig:combined_PAD}}
\end{figure*}

Figure~\ref{fig:combined_PAD}a shows the electron event observed by \textit{SolO}/EPT in the 40--50~keV energy channel. The top panel, which shows the intensities as measured by the four viewing directions of EPT, reveals a strongly anisotropic event starting at $\sim$6:40~UT followed by an isotropic phase starting around 7:15~UT. The second panel shows the pitch-angle coverage of EPT's four viewing directions, which is ideal and exceptionally stable over the course of the event. The good pitch-angle coverage allows not only to determine the strong first order parallel anisotropy (shown in the bottom panel) in the early phase of the event, but also to verify that the later isotropic phase is real and not caused by poor-pitch angle coverage.

Figure~\ref{fig:combined_PAD}b shows the electron event at 1.1--2.4~MeV measured by \textit{SolO}/HET, which provides the same four viewing directions, allowing us to determine the pitch-angle distribution also at MeV energies. The initial phase of the MeV-electron event shows a significantly smaller anisotropy. However, the time series still reveals a two-component picture with the later, almost isotropic phase starting like the one observed by EPT around 7:15~UT. A smaller anisotropy for higher electron energies is expected because of the energy-dependence on pitch-angle scattering, which is stronger for higher energy electrons than for the lower energy ones \cite[][]{Droege2000, Agueda2014, Strauss2020}.
However, in contrast to the low energy electrons (Fig.~\ref{fig:combined_PAD}a), the intensity level of the two components is rather similar at MeV energies, with the later component being even slightly more intense. At 40--50~keV (Fig.~\ref{fig:combined_PAD}a), the first, anisotropic component is much more intense than the second, isotropic component, indicating a softer spectrum of the first component. This is confirmed by analysing the peak spectra of both components (see Section~\ref{Subsec:spectra}).

Figure~\ref{fig:combined_PAD}c shows the 45--55~keV electron event observed in the four viewing directions of \textit{STEREO-A}/SEPT. Because the spacecraft was put upside down after its superior solar conjunction in 2014, the SEPT sun and anti-sun sectors do not point anymore along a nominal Parker spiral field line but perpendicular to it. This causes often non-ideal pitch-angle coverage, which is, however, not too bad during the electron event of 9 October 2021, as shown by the 2nd and 3rd panels of Fig.~\ref{fig:combined_PAD}c. The onset of the event is observed by the sun sector of SEPT. However, during the rise phase of the event, the magnetic field direction changes at $\sim$7~UT, so that the anti-sunward propagating beam is then observed best in the anti-sun telescope. We note that the pointing of the SEPT instrument has changed after the superior conjunction in 2015, since when the spacecraft was put upside down. Therefore, the north and south telescopes are swapped and the sun and anti-sun telescopes point perpendicular to the nominal Parker spiral \citep[cf. Fig. 8 in ][]{Badman2022}
At the time of the onset the pitch-angle coverage is ideal showing a smaller anisotropy than at {\it SolO}. The rise phase of the event also shows a smaller anisotropy than observed by {\it SolO}, with no indication of two distinct components. However, this could be potentially masked by the non-ideal pitch-angle coverage during the rising phase of the event. Unfortunately, \textit{STEREO-A}/HET provides only one viewing direction, which is aligned with the pointing of SEPT-sun so that we cannot infer the anisotropy at MeV energies for \textit{STEREO-A}.

Sectored electron observations by \textit{Wind}/3DP are shown in Fig.~\ref{fig:combined_PAD}d with a longer time period than shown in the other panels. The event is much more gradual and less intense. However, the initial phase shows some anisotropy, followed by a second step with less anisotropy. At a first glance, this might appear like the same two-component picture as observed by \textit{SolO}, but the time periods are significantly different. While the first, anisotropic component lasts about 30~min at \textit{SolO}, it has a duration of about five hours at \textit{Wind} so that a one-to-one correspondence is unlikely. The second increase coincides with the crossing of a magnetic sector boundary and is therefore likely related with a change of magnetic connectivity rather than with a new injection at the Sun.

\begin{figure*}[h]
\centering
\includegraphics[width= 0.7\textwidth]{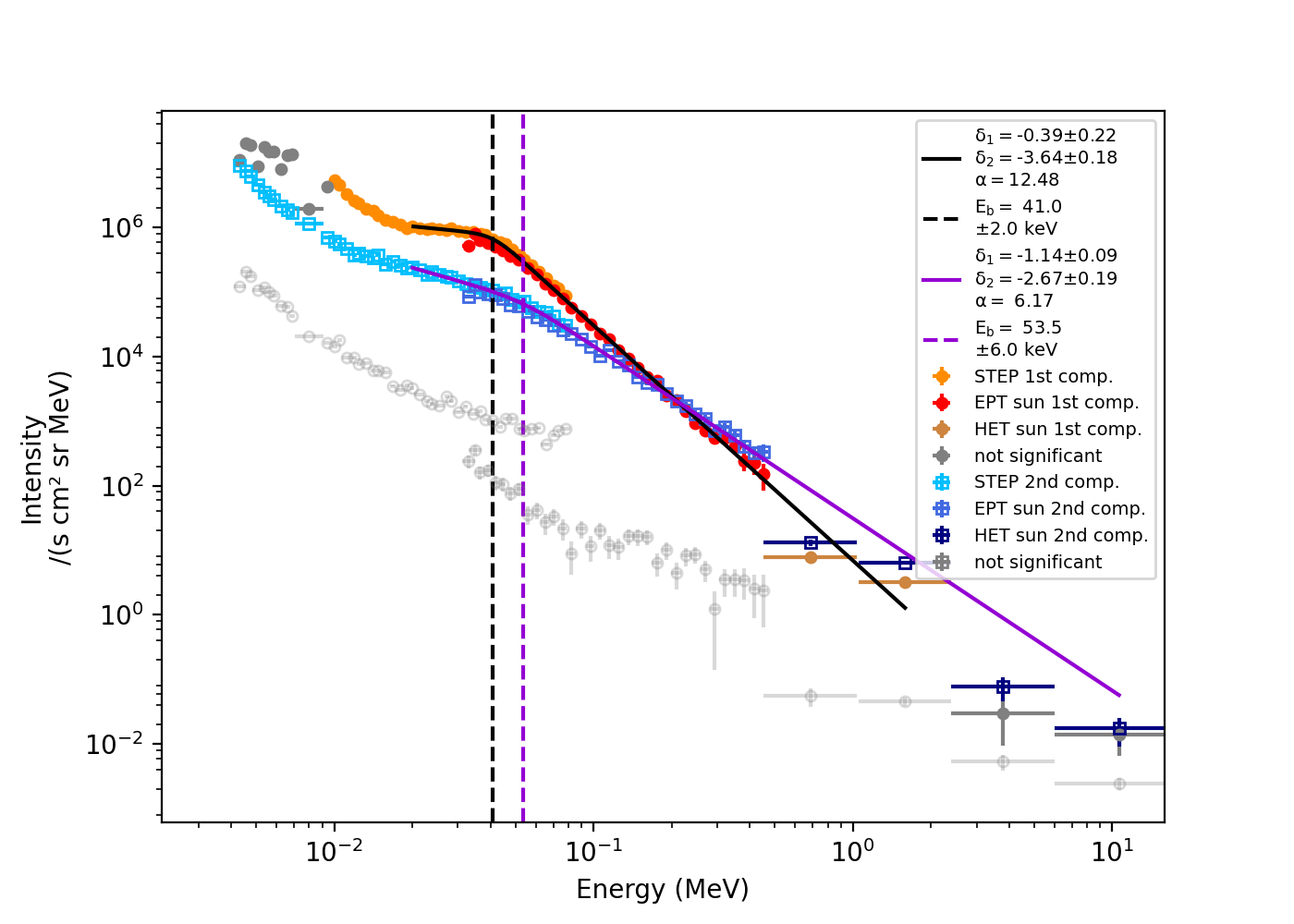}
\caption{Electron peak intensity energy spectra determined for the two SEP components of the electron event observed by {\it SolO}.  \label{fig:combined_EPD_spec}}
\end{figure*}

\subsubsection{Energy spectra observed by SolO} \label{Subsec:spectra}
{\it SolO} is the only spacecraft that clearly observes a two component electron event as described above. Figure~\ref{fig:combined_EPD_spec} shows peak electron spectra observed by the three EPD instruments STEP (orange), EPT (red), and HET (brown). For EPT and HET, we use the sunward-looking telescopes that align with the viewing direction of STEP and cover the usual incoming direction of SEPs. The figure shows a spectrum for each of the two components, with circles (squares) denoting the first (second) component. The light-grey points represent the pre-event background spectrum that has been subtracted from each of the two spectra. Points with dark-grey colour represent energy bins that did not show a significant increase above the background level (HET) or which were contaminated by ions (STEP). 
Both spectra have been fit with a broken power law using the methods described in \cite{Dresing2020, Strauss2020}. 
The resulting spectral indices below and above the break as well as the break energy are provided in the figure legend. We only use part of the STEP energy channels in the fit to avoid fitting the upturn in the very low energy range observed by STEP, which could be caused by mixing with another low-energy event. We also exclude the first energy channel of HET from the fit because of instrumental effects that lead to too low intensity measurement, which has been fixed in a patch uploaded to the spacecraft only after this event. 
Compared to the first component, which has a spectral index above the break energy of ($\delta_2=-3.64\pm0.18$), the spectrum of the second component is clearly harder ($\delta_2=-2.67\pm0.19$).
At energies $\gtrsim 200$~keV, the spectra even intersect, meaning that the second component is more intense than the first one at these energies. This suggests that the second component is formed by a distinct particle injection, rather than being a part of the first component. Furthermore, the significantly harder spectrum of the second component suggests a more efficient acceleration process.

\subsubsection{Release times} \label{Subsec:releasetimes}

The electron event is well observed at \textit{SolO}, especially in the near-relativistic range covered by EPT. We determine the onset time at the spacecraft of each energy channel using a statistical quality control scheme that is designed to decide when the quality of the monitored variable changed from being in control to out of control. There are many different quality control schemes, and cumulative sum (CUSUM) schemes are used in many industries due to their capability to give an early warning of changes in the monitored process \citep[][]{page54}. A traditional CUSUM method assumes that the variable in question is normally distributed, and in the case of the monitored variable having a Poisson distribution, the quality control scheme used should be the Poisson-CUSUM \citep[][]{lucas85, Huttunen2005}, which we apply here for the determination of the onset times. 

For the parameters of the method we use the mean and standard deviation of intensity measurements in the chosen pre-event background. We average the original measurements from 1 second resolution to 30 second resolution, in order to minimise the effects of counting noise. Furthermore, to avert false positives in the determination of onset times, we demand that at least 60 consecutive out-of-control signals must be counted by the method before identifying the onset of the event as the first of these signals. This limit corresponds to 30 minutes of intensity measurements exceeding the threshold of Poisson-CUSUM method.

\begin{figure}[ht]
\centering
\includegraphics[width= 0.49\textwidth]{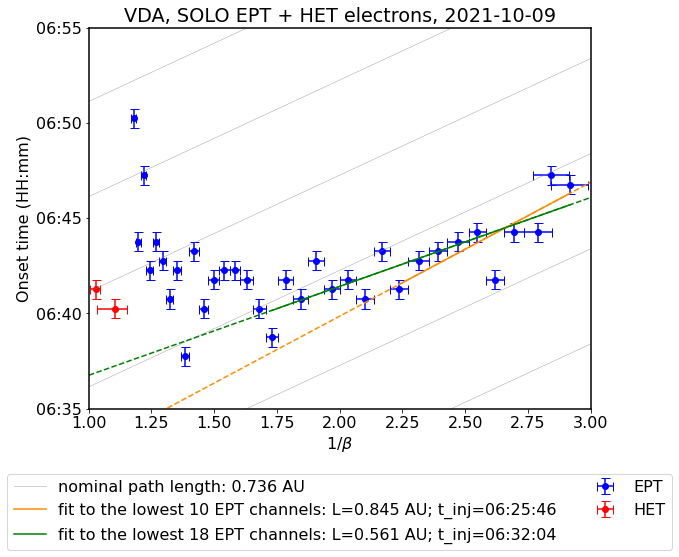}
\caption{Velocity dispersion analysis based on onset times of \textit{SolO} EPT and HET electron channels. The horizontal axis shows the inverse of the average unitless speed of electrons as observed in each channel. The vertical axis presents the determined onset time for each energy channel. Onset times observed by EPT (HET) are marked in blue (red). Horizontal error bars represent the width of the energy channels, and vertical error bars represent the time resolution used to determine the onsets (30 seconds). The orange and green lines are linear fits to the lowest 10 and 18 energy channels of EPT, respectively. The grey lines represent the slope corresponding to a path length matching the nominal Parker spiral length at \textit{SolO}'s radial distance. \label{fig:vda_solo}}
\end{figure}

Using these onset times, we perform a velocity dispersion analysis \cite[VDA, e.g.,][]{Vainio13} to infer the common solar injection time of the observed electrons. Figure~\ref{fig:vda_solo} shows the onset times as a function of the inverse unit-less speed ($1 / \beta = c/v$) of the electrons. Blue symbols correspond to the onset times observed in the different energy channels of EPT, which is capable of observing electrons in the energy range 31.2--471.4~keV, and red symbols mark HET observations, which measures electrons in the MeV range. For some channels the onset could not be resolved, which is why there are fewer data points in the plot than there are total channels in EPT and HET combined. 

The VDA method assumes that particles of all energies were injected at the Sun into the interplanetary space at the same time and that their propagation path length is identical. Consequently, the onset times as a function of inverse beta should align themselves to a line with a rising slope. This slope then allows to determine the length of the path travelled by the electrons, and the intersection with the vertical time axis tells the time of the common particle injection. The grey lines in the background of Fig.~\ref{fig:vda_solo} represent slopes that one would get with the nominal Parker spiral path length of 0.736~AU, assuming a solar wind speed of $v_{sw} = 360~\mathrm{km/s}$ and taking into account \textit{SolO}'s heliocentric distance.
As can be seen in Fig.~\ref{fig:vda_solo}, the onset times do not follow this trend, nor do they form a straight line. Instead, the higher energy channels of EPT show more and more delayed onsets. 

Above an energy of about 142~keV, it becomes unreasonable to include further data points into the VDA, since the general trend from that energy channel upward are more and more delayed onsets. We therefore perform a fit only to the first 18 EPT lowest energy channels (shown by the green line). However, this results in a path length of L=0.56~AU, which is nonphysical (i.e., too short) since the spacecraft's radial distance is 0.68~AU. We therefore apply another fit only to the 10 lowest energy channels (31.2--62.7~keV), which is shown by the orange line. This fit yields a reasonable path length of $L=0.845 \pm 0.380~\mathrm{AU}$ and an inferred injection time at the Sun for electrons of these energies at 6:25:46~UT $\pm$ 476~s.

\begin{figure*}[ht]
\centering
\includegraphics[width=0.85\textwidth]{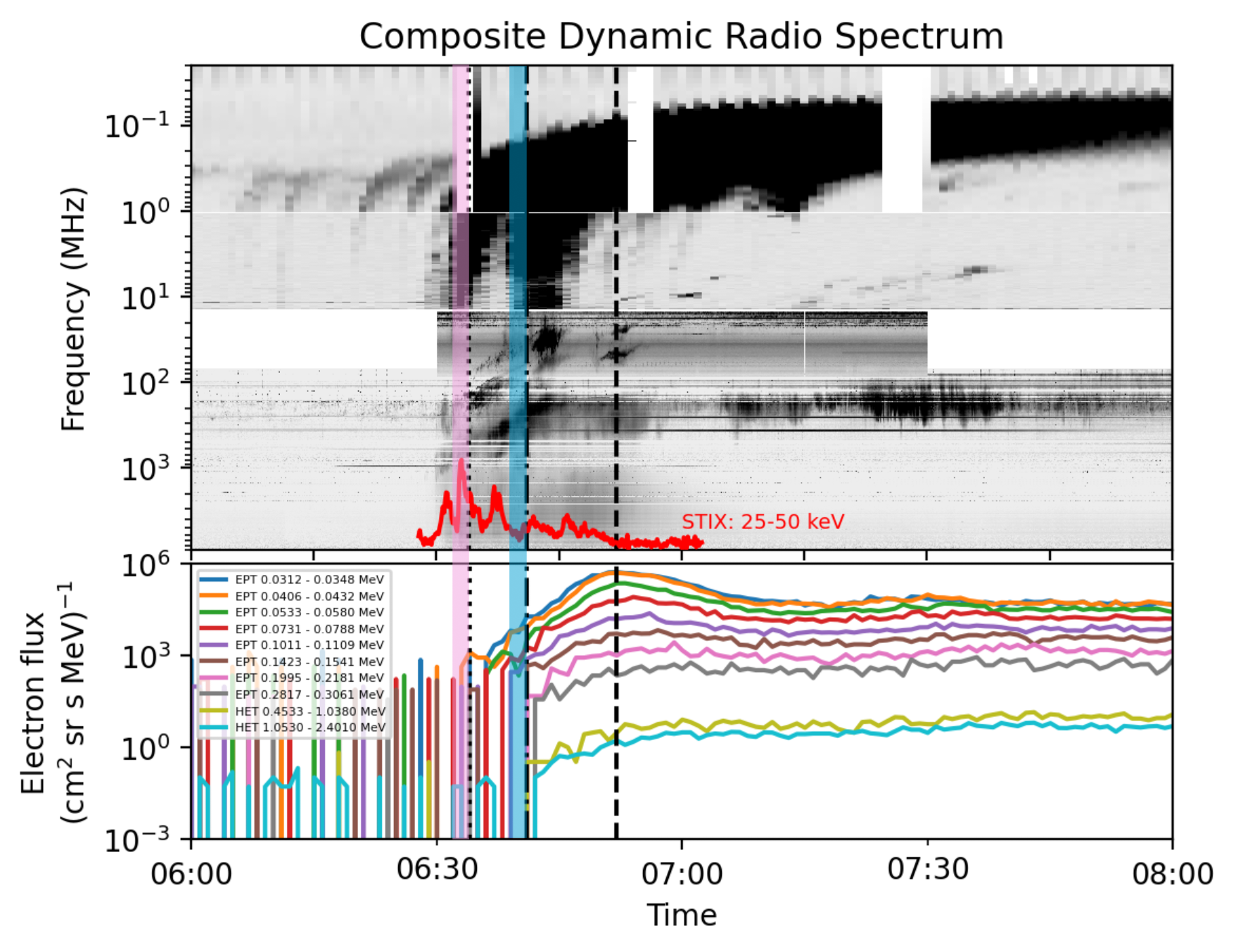}
\caption{Composite spectrum of all available radio observations from the Earth vantage point (9 GHz - 10 kHz) together with STIX nonthermal (25 \-- 50 keV) count rates (\textit{top panel}), and the in-situ electrons observed at \textit{SolO} (\textit{bottom panel}). The \textit{SolO} electron observations from the sunward telescope are time-shifted using the nominal path length and corrected for the travel time of light to make them comparable to the radio observations. The inferred injection time of the low-energy (high-energy) electrons are marked with the dotted (dot-dashed) line. With the pink (blue) shaded area we mark the release times using path length values larger than the nominal. For $\sim$0.85~AU, the earliest release of the low-energy electrons is near the start of the first type III. The dashed line indicates the peak of the electron intensity for the $\sim$55~keV electron channel. \label{fig:Radio_and_Electrons} }
\end{figure*}

Like for the high energy channels of EPT, the onset times of MeV electrons as observed by \textit{SolO}/HET (red points in Fig.~\ref{fig:vda_solo}) are also systematically delayed and do not fit the velocity dispersion trend as indicated by either fit. This suggests a delayed solar injection with respect to the lower energy electrons. We apply a time shift analysis \cite[TSA, e.g.,][]{Vainio13} to the onset time of the 1.05--2.41~MeV energy channel using a nominal Parker spiral length of $L=0.736~\mathrm{AU}$ and a solar wind speed of 360~km/s in order to infer the solar injection time of these electrons. This channel was selected because it has the clearest and most resolvable onset. We obtain an injection time of 06:35:56~UT at the Sun, which is about 10 minutes later than that for the $\sim$30--60~keV energy electrons using TSA. For same low energy channel, using the path length determined by the VDA ($L=0.845~\mathrm{AU}$), we only obtain a one minute earlier injection time. This strongly suggests the delayed solar injection of the high-energy electrons compared to the low-energy electrons. An application of the TSA is also presented in Fig.~\ref{fig:Radio_and_Electrons} for a number of energy channels assuming a path length larger than normal ($L=0.85~\mathrm{AU}$).


We also performed a VDA analysis for \textit{STEREO-A}. From the onset times determined from the \textit{STEREO-A}/SEPT electron channels, a trend similar to \textit{SolO}/EPT arises: The onset times in the highest channels, namely those above 195~keV, are consistently more delayed. On the other hand, a fit to the lower energy channels in the range 45--195~keV yields a nonphysical path length of $L=0.53~\mathrm{AU}$. If we instead choose the onset time as the moment when the measured electron intensity reaches 1\% of the event peak intensity, then a fit to the same energy range yields a longer, but still nonphysical path length of $L=0.72~\mathrm{AU}$. If we instead use TSA on onset time of the 45--55~keV channel to infer the solar injection time, we get $t_{j} =$ 06:28:50, assuming the nominal Parker spiral arm for a solar wind speed of 390~km/s. Similarly, for 0.7--1.4 MeV electrons  measured by STEREO/HET we obtain a solar release time of $t_{j} =$ 06:47 UT, almost 20 minutes later than that of the low-energy electrons. Table~\ref{table:onset_times} summarises the results of the VDA and TSA. It shows the VDA result for {\it SolO} only for the more reasonable VDA fit and also results of similar TSA analyses done for \textit{STEREO-A} SEPT and HET (not shown here in a figure).

\begin{table*}
\caption{Electron onset and inferred solar injection times.}             
\label{table:onset_times}      
\centering          
\begin{tabular}{l l l l l r}     
Observer /  & Energy  & Onset time &  Inferred injection  &  Path length & Method\\
Instrument & (keV) & at spacecraft (UT) & time at Sun (UT) & (AU) & \\
\hline
SolO / EPT & 31.2 - 62.7 & - & 6:25:46 $\pm$ 0:07:56 & 0.845 $\pm$ 0.380 &VDA \\
SolO / HET & 1053 - 2410 & 6:41:15 $\pm$ 0:00:30 & 6:34:56 $\pm$ 0:00:30 & 0.736 &  TSA \\ 
STEREO-A / SEPT & 45.0 - 55.0 & 6:50:58 $\pm$ 0:01:00 & 6:28:50 $\pm$ 0:01:00 & $1.096 \pm 0.1$ & TSA \\
STEREO-A / HET & 700 - 1400  & 6:57:00 $\pm$ 0:01:00 & 6:47:18 $\pm$ 0:01:00 & $1.096 \pm 0.1$ & TSA \\ 
\hline                    
\end{tabular}
\end{table*}

\section{Discussion}

\begin{figure*}[ht]
\centering
\includegraphics[width= 0.99\textwidth]{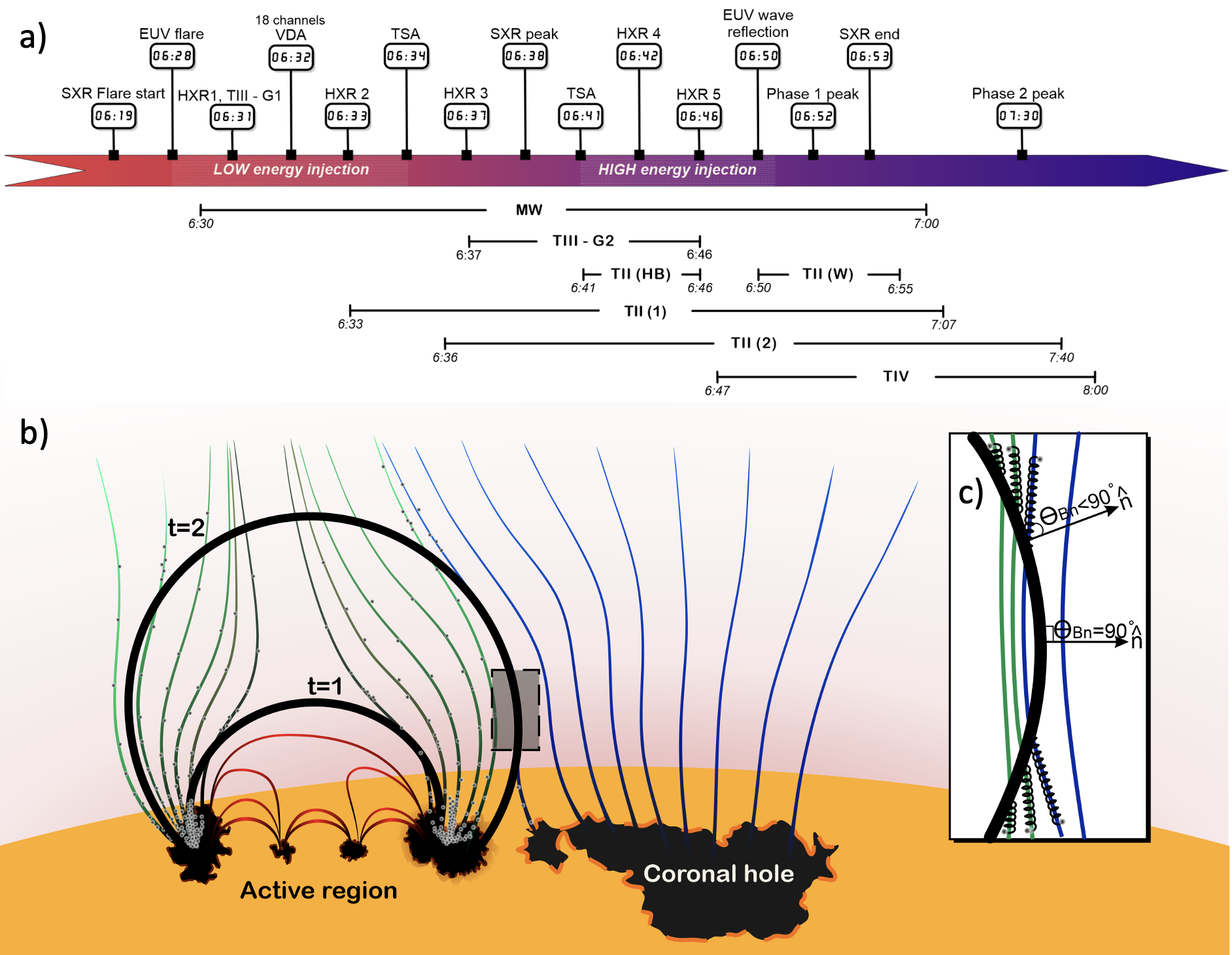}
\caption{Events during the first phase leading to the first in-situ component. (a) Timeline of the features associated with the two different electron injections during the first phase. The exact times of the observational features are marked on the timeline bar. The gradient on the bar is red to blue, indicating the transition between the low and high energy releases close to the Sun. The long-lasting radio features are marked below the timeline. (b) Cartoon representation of the first phase. The coronal shock wave during the low-energy release (t=1)  and high-energy release (t=2) are represented by the black arcs. The open field lines from the periphery of the active region are represented by the green lines, while those from the adjacent coronal hole are represented by blue lines. The closed field lines in the AR are denoted by th red coloured lines. The abundance of energetic electrons are denoted by the small grey dots. The transparent grey rectangle is zoomed into in panel (c) and shows the shock wave's lateral expansion into nearby open magnetic field lines. The shock normal is denoted by the arrow and the $\hat{n}$. The perpendicular and near-perpendicular geometry is represented by $\theta_{Bn} \approx 90^{\circ}$ and $\theta_{Bn} < 90^{\circ}$. The electrons accelerated at the shock front on either side of the $\theta_{Bn} \approx 90^{\circ}$ part of the shock wave are marked by the spirals along the magnetic field lines. More details provided in the text. \label{fig:timeline}}
\end{figure*}

In this study, we have analysed an event on 9 October 2021, which was associated with an electron rich  event in-situ and was also observed by multiple wavelength remote sensing observations. In our analysis, we have considered observations from X-rays, microwaves, radio waves, EUV, WL, and in-situ measurements to understand the solar sources of this electron event. We also utilised the full capabilities of the Solar Orbiter (\textit{SolO}) mission. The many features of the event are summarised with the use of a timeline inferred from both remote sensing and in-situ observations (Fig.~\ref{fig:timeline}).

The in-situ observations of energetic electrons by \textit{SolO} show distinct phases of electron energization that are indistinguishable at other spacecraft. More specifically, the energetic electrons observed by \textit{SolO} show two increases in the electron intensity within two hours. Additionally, the electrons' anisotropy shows clearly that there are two distinct phases (Sec.~\ref{subsec:electron anisotropy}), namely, a first anisotropic phase, which is observed immediately after the onset of electrons in the spacecraft, and a second mostly isotropic phase that is observed after the first peak in the in-situ electron intensities. Throughout the Discussion, we will term these two phases observed by \textit{SolO} as "Phase 1" and "Phase 2". A further analysis of "Phase 1" and more specifically the electron release times shows that the low- and high-energy electrons are released at different times.

We also note that such a distinction of two phases was not recorded for the electrons observed by \textit{STEREO-A} and \textit{Wind}. The two peaks in the electron intensity that were observed at \textit{SolO} were also not seen by \textit{STEREO-A} and \textit{Wind}. In the following subsections, we will discuss the two different phases as characterised by the anisotropy that was observed by \textit{SolO} and their possible acceleration mechanisms at or near the Sun.

\subsection{Phase 1} \label{subsec:phase_1}

The first phase extends from immediately after the first electrons arriving at \textit{SolO}, until 07:14~UT, when the initially peaked electron anisotropy vanishes. During the first phase, we inferred that the low energy electrons ($\lesssim 142$~kV) were released $\sim$5--10 minutes earlier than the more energetic electrons (Sec.~\ref{Subsec:releasetimes}). Given this time difference, we can distinguish several solar features associated uniquely to each release. HXR observations show distinct pulses around the electron release and radio observations show the escape of electrons to open field lines and the formation of multiple strong shock regions during this phase.

An anisotropic early phase of the event is not only observed at \textit{SolO}, but also at other observing spacecraft, namely, \textit{STEREO-A} and \textit{Wind} (Fig.~\ref{fig:combined_PAD}). Such a consistent increase in anisotropy among multiple observers may indicate a common acceleration/release process. It is unclear whether the same event was observed at \textit{Wind}, since the electrons arrive an hour after their arrival at \textit{SolO} or \textit{STEREO-A}.  


The intensity-time profiles of the in-situ electrons recorded by \textit{SolO} during this phase were impulsive, highly anisotropic from the sunward direction, and exhibited a typical power-law energy spectrum with a hard-soft spectral index (Fig.~\ref{fig:combined_EPD_spec}). This broken power-law energy spectrum of the in-situ electrons has been previously suggested to be due to beam propagation through inhomogeneous plasma \citep[][]{Krucker08, Kontar09}. Additionally, high anisotropy for this phase was generally observed by all spacecraft within the 50$^{\circ}$ heliolongitudinal range, which suggests that the electron injection took place low in the corona and at regions well connected to most of the observers.


\subsubsection*{Low-energy electrons}

The inferred injection time of the low-energy electrons is $\sim$06:32~UT $\pm$7:30 min based on the VDA and $\sim$06:34~UT using TSA at $\sim$42~keV electrons (both converted to the spacecraft time frame). A number of flare- and shock-associated features were observed during this time. The HXR observations presented in Sec. \ref{subSec:Xray} show distinct pulses around the electron release time, which were also observed in microwaves along with a characteristic diffuse gyro-resonant component indicating the presence of trapped electrons. 
The analysis of the HXR photon spectrum also shows that electrons were routinely accelerated to energies of $\sim$10\,--\,100~keV, suggesting a flare-related contribution to the low-energy electrons observed in-situ.

The inferred injection time of the low-energy electrons seems to be associated with the first two HXR pulses and the TIII-G1. The TIII-G1 was probably generated by electron beams released from the flare site. 
Space-borne radio wave observations of the TIII-G1 reveal that the source electron beam was propagating at roughly  -50$^{\circ}$ heliolongitude (Fig. \ref{fig:space_obs_all}), that is, away from the Parker spirals connecting \textit{SolO}, \textit{STEREO-A}, and \textit{Wind}. 
According to our analysis (Appendix~\ref{Sec:appendix_radio}), the in-situ Langmuir waves observed at \textit{STEREO-A} and \textit{SolO} were unlikely to be generated by the passing of TIII-G1 (Fig. \ref{fig:space_obs_all}). 

According to Fig.~\ref{fig:Radio_and_Electrons}, another physical phenomenon was present around the time of the inferred low-energy electron release and that was the coronal shock wave. The first signatures of TII(1a) were observed around 06:33~UT, indicating that the shock was able to efficiently accelerate electrons after the first HXR pulse. Around the same time, the EUV wave was formed and started its evolution away from the active region, mostly to the southwest direction. According to the first-order analysis of the shock wave (type II association presented in Appendix~\ref{Sec:appendix_shock}), the TII(1a) originated most likely from the extended flank regions of the coronal shock wave. Furthermore, as shown in Fig.~\ref{fig:EUV_wave} and Sec.~\ref{subSec:Shockwave}, open magnetic field lines indicate a good magnetic connectivity to both \textit{SolO} and \textit{STEREO-A} at the periphery of the AR, where the shock is progressively connected. 

In summary, we have found that the low-energy electrons were accelerated mainly by a flare-related process consistent with the first two HXR pulses. Fig.~\ref{fig:nlfff} shows that the flare occurred in the western part of the AR. There, open field lines are present only in the vicinity of the strong positive polarity in the south. During the first (shown in panel a and b) and second HXR peak, a footpoint source is located close to this area. In contrast, in the later peaks the eastern footpoint has shifted to a more northern location. We conclude that this is consistent with the notion that the energetic electrons observed in-situ in phase~1 have been accelerated (at least partly) in the flare and were injected onto open magnetic field lines connection to IP space.

We also found that a relative contribution from the shock wave is difficult to be excluded, since the inferred injection and the start of type II are very close. The shock wave possibly facilitated the low-energy electron release to the open magnetic field lines connecting to different observers, since the EUV wave reached the well-connected field lines (to \textit{SolO} and \textit{STEREO-A}) around 06:35~UT. During this phase, the shock geometry at the field lines is expected to be mostly quasi-perpendicular at its flanks as it interacts with the open field lines present close to the active region. This may allow the shock wave to accelerate electrons rapidly along open magnetic field lines via the SDA mechanism. This scenario is illustrated in Fig.~\ref{fig:timeline}b where the shock wave during this time is represented by the black arc at t=1. The thermal and suprathermal electrons which are present in abundance in the vicinity of the active region can be accelerated to energies in the range of 40--80 keV through a fast-Fermi mechanism. As explained earlier, given the right coronal and shock conditions (upstream electron density, magnetic compression ratio, and shock geometry), SDA can be a potent accelerator of energetic electrons \citep[][]{Leroy84, KraussV1989a, Mann09, Warmuth09b}. 

\subsubsection*{High energy electrons}

The inferred injection times of electrons exceeding an energy of ${\sim}100$~keV (using VDA and TSA) show a delay of 5 to 10 minutes, compared to that of the low-energy electrons (Sec.~\ref{Subsec:releasetimes}). This puts it amidst several observed eruptive features and potential acceleration scenarios. At the time of the electron release, the TIII-G2, the 4th HXR pulse, and the TII(HB) were observed. We note that there was also a restructuring of magnetic footpoints near the flaring region minutes before the injection, during the 3rd HXR pulse (Fig.~\ref{fig:stix_images} and Sec.~\ref{subSec:Xray}).


The TIII-G2 was observed in association with the 4th HXR pulse, which was also one of the two lowest intensity pulses, implying that there was a decrease in the number of available energetic electrons (Fig.~\ref{fig:stix_goes_spectral}f). The associated TIII-G2, however, was rather bright (dense electron beams generating the type III) and with a prolonged duration. This often signifies an extended period of electron release into interplanetary space. It is not clear if parts of the TIII-G2 emanated from the type II or not. The TIII-G2 was observed by all space-borne radio observatories and, in particular, by \textit{STEREO-A} and \textit{SolO} that also observed in-situ Langmuir waves. The results from the direction-finding analysis of the TIII-G2 presented in Appendix~\ref{Sec:appendix_radio} reveal that the TIII-G2 propagated approximately in the -15$^{\circ}$ heliolongitude (Fig.~\ref{fig:space_obs_all}), which was towards both \textit{SolO} and \textit{STEREO-A}. The radio analysis suggests a good magnetic connection between the source of the TIII-G2 and the two spacecraft.

During this phase, energetic electrons in the range of 50--100 keV were still found in abundance in the flaring region (Sec.~\ref{subSec:Xray}). Additionally, important shock-associated phenomena were also observed, namely, TII(2a), TII(HB), and TII(W) at 06:36~UT, 06:41~UT, and 06:50~UT, respectively. The presence of herringbone (HB) structures indicate that electron beams were accelerated by a near-perpendicular shock wave ($85^{\circ} < \theta_{Bn} < 89.9^{\circ}$) via the SDA mechanism \citep[][]{Mann05}. 
Such a near-perpendicular shock geometry is expected in the low corona, which is dominated by closed magnetic field lines and a strong lateral expansion of the shock wave \citep[][]{Kouloumvakos21}. According to the analysis performed in Sec.~\ref{subSec:Shockwave}, it is understood that the TII(HB) and TII(W) features were observed concurrently with the EUV wave mirroring off a magnetic null, which is formed in the south-west periphery of the AR close to the south-west coronal hole. These two features mark a phase of efficient electron acceleration from the shock, which is probably responsible for the injection of the high-energy electrons during the first phase. Meanwhile, the TII(2a) was observed in metric wavelengths and according to the kinematic analysis presented in the Appendix~\ref{Sec:appendix_shock}, the source of the TII(2a) is at the flank regions of the shock wave.

A critical feature of this period was the possible interaction between the electron beams generating the TIII-G2 and part of the shock wave generating TII(HB) leading to a re-acceleration of the incident electrons. Based on the interplanetary directivity analysis of TIII-G2, we have shown that the source of the electron beams were well-connected to both \textit{SolO} and \textit{STEREO-A} (Fig.~\ref{fig:space_obs_all}). Additionally, during the same period, the TIV continuum observed was probably generated by electrons trapped within the flux rope of the expanding CME. It is possible that some of the trapped energetic particles may have escaped during interactions between the CME and the ambient magnetic field lines \citep[][]{Petrosian16, Klein22}. In both circumstances, an interaction with the shock wave could have accelerated the incident electrons to higher energies. Electrons with large enough pitch angles are readily reflected by a near-perpendicular shock wave, gaining maximum energies of up to 13-fold their initial energy in the process \citep[for a shock wave with magnetic compression 4,][]{Leroy84, Ball2001}. 
This mechanism may have increased the energy of a fraction (depends on $\theta_{Bn}$) of the incident $c/3$ electrons that generated the TIII-G2 to near-relativistic energies.
Considering that electrons in the range of 50--100 keV were also abundant during this period from, e.g., the HXR pulse, they may be further accelerated to relativistic energies by interacting with the coronal shock wave.


In conclusion, we find that during the release of high-energy electrons, a number of processes were observed concurrently. We distinguished between the different features and their associated mechanisms and found that the acceleration most likely happened in the corona close to the Sun. Observations suggest that the main acceleration probably took place at the southern flank of the shock wave, where an interaction of the shock with a coronal streamer is observed. In these regions of interaction the shock wave geometry can be nearly perpendicular and electrons can accelerate to high energies efficiently. A cartoon illustrating such a scenario is presented in Fig.~\ref{fig:timeline}b,c, where the shock wave (represented by the black arc at t=2) continues to interact with open field lines at its flanks. The presence of nonthermal electrons probably eases the acceleration of the electrons to mildly-relativistic and relativistic energies; at least for a fraction of the incident electrons. 

A contribution from a flare-related acceleration process cannot be excluded during this high-energy electron release phase, however, the Suprathermal Ion Spectrograph (SIS) on-board \textit{SolO} showed no particular enhancement of flare accelerating particles such as $^3$He ions. The presence of $^3$He ions would indicate a direct contribution of flare accelerated particles to the particle population measured in-situ during the event. This probably suggests that the shock wave had an important role in the acceleration and release of this high-energy electron component. The shock geometry which in some cases have been shown to be predominantly quasi-perpendicular in the low corona \citep[e.g.,][]{Kouloumvakos21}, can enable the shock to accelerate electrons up to 13-fold their incident energy based on their pitch angles through an SDA mechanism \citep[][and references therein]{Leroy84, Ball2001}.

\subsection{Phase 2} \label{subsec:phase_2}



As observed in Fig.~\ref{fig:Radio_and_Electrons}, a second peak was observed around 7:30~UT in the electron time profiles recorded by \textit{SolO}. This increase in intensity appears after the anisotropic first phase. In this second phase, the electrons showed very little anisotropy (Fig.~\ref{fig:combined_PAD}), which is usually believed to be due to transport related effects. 
It was not possible to determine the release of these electrons, however, the peak intensities were delayed by $\approx$50 minutes after the start of the SXR flare (Fig.~\ref{fig:timeline}). This phase can be characterised by the gradual increase in high-energy electron intensities. An interesting aspect of this phase was the increase in the intensities of energetic electrons exceeding 300 keV, which was only observed by \textit{SolO} and not by other spacecraft. Additionally, the high-energy electrons exhibited a harder spectrum than during the first phase as observed by \textit{SolO} (Fig.~\ref{fig:combined_EPD_spec}). A hard spectrum for delayed and gradual electron events has previously been attributed to transport-related effects \cite[][]{Strauss2020}. 


In the case of \textit{STEREO-A}, we believe the spacecraft sampled electrons that were accelerated mostly during the first phase and were delayed due to transport related effects. This could explain the anisotropy of the electrons observed by \textit{STEREO-A}, which was lower than at \textit{SolO}. This would suggest that \textit{SolO} was located conveniently at a region where the electrons were largely unaffected by transport and did not undergo strong pitch-angle scattering. The other spacecraft located elsewhere might have observed electrons from both the first phase and a transport-affected second phase. 


The second phase occurred after the impulsive flare phase, that is, when the HXR flux had returned to nominal pre-flare levels. It is therefore probable that the electrons of phase 2 were accelerated at the CME/shock instead of during the flare. In radio wavelengths, the only features remaining in the radio spectrum are a hectometric type II burst and metric TIV pulsations. At the same time, the WL shock wave reached its peak velocity of 1400~km~s$^{-1}$ (Fig.~\ref{fig:shock_kinematics}) at a radial distance of 7~$R_{\odot}$. The presence of a long-lasting patchy type II radio burst in the hectometer wavelength, namely TII(1b) and TII(2b), confirms that shock electron acceleration took place during this phase. From the analysis of Sec.~\ref{subSec:Shockwave}, it is understood that the shock underwent large-scale deformations due to the presence of coronal magnetic structures, namely, coronal holes and streamers, and it also continued to expand in regions confined by different stream interactions (Sec.~\ref{subSec:Shockwave}). An analysis of the shock kinematics (Appendix~\ref{Sec:appendix_shock}) and the comparison with the radio TII kinematics suggests that TII(2b) was most likely generated in the flank regions of the WL shock wave observed during this second phase. 

Despite the fact that SDA can be a potent accelerator of energetic electrons in regions with quasi-perpendicular shock geometry, it is not fully clear if the shock geometry in the high corona is such. 
It is possible that small scale deformations such as cavities and ripples may provide the locally quasi-perpendicular geometry. However, in these places the shock can be corrugated, leading to the short and efficient acceleration of electrons trapped within them through SDA mechanism \citep[][]{Bale99}. On the other hand, DSA can also be an efficient acceleration mechanism if the electrons are trapped efficiently at the shock region. In this case, pre-existing magnetic field turbulence and low frequency waves generated upstream by specularly reflected ions \citep[about 25\% of the thermal ions are reflected; ][]{Krasnoselskikh91, Gedalin96, Balikhin22} could have played a crucial role in trapping electrons for an extended period. Ambient magnetic field turbulence, which is ubiquitous in interplanetary plasma, may act as magnetic scattering centres for electrons \citep[][]{Tsytovich66}. 
The electrons trapped within turbulent fields or deformed and corrugated shock regions may diverge from the thermal Maxwellian \citep[][]{Maksimovic97} and provide a seed population for DSA. The energy gain of these electrons can be in the order of 10s of keV \citep[resonance with whistler waves at upper and lower-hybrid frequency; ][]{Vaisberg83, Galeev84, Krasnoselskikh85} to several MeVs and tens of MeVs \citep[Alfven wave resonance; ][]{Tsytovich73, Bell78, Kirk01} depending on how efficiently they are trapped near the shock.

The fact that this second phase was only observed by \textit{SolO} indicates that the conditions for the shock scenario were provided only within a confined region of the interplanetary shock wave. The confined region could have been formed when the WL shock wave interacted with the streamers and propagated close to the fast-wind region. The electrons which were accelerated through a pitch-angle diffusion process under the aforementioned conditions can then escape the shock. While even small amounts of perpendicular diffusion present in the SIR can spread the particles to large heliolongitudes, we do not observe the second phase in a spacecraft other than \textit{SolO}. 

The intensity-time profiles of the SEP event at \textit{SolO} might be influenced by the arrival of a stream interaction region (SIR) during the SEP onset time. As discussed by \citet{Lario22}, the high-speed stream driving the SIR was previously observed by \textit{STEREO-A} and later also by near-Earth spacecraft, where it strongly affected the observed energetic ion intensity-time profiles. At \textit{SolO}, the SIR and its associated magnetic compression might have acted as a magnetic mirror, reflecting some of the electrons back towards the inner heliosphere and the CME-driven shock wave where the electrons may subsequently accelerate to higher energies. 


\section{Conclusions} \label{Conclusions}



The main conclusions of this study are as follows:

\begin{itemize}
    \item [\textbullet] We were able to identify two distinct SEP event phases in-situ, that show different anisotropies and are most-likely related to different acceleration phases in the solar corona. 

    \item [\textbullet] We also found a hardening spectra from the first to the second phase for mildly-relativistic and relativistic electrons. This finding further prompts the existence of two different acceleration mechanisms. 

    \item [\textbullet] We have found the presence of two separate injections during the first phase. The low-energy and high-energy electron components were injected at different times and can be attributed to different episodes of acceleration. 

    \item [\textbullet] We found a mix of flare and shock acceleration mechanisms for the low energy electrons, while the high energy electron acceleration was mostly a shock-associated phenomenon in a highly structured corona.

    \item [\textbullet] We have also found that the prolonged nature of the second phase was due to the particles being accelerated and transported diffusively within a compression region, where the said particles are also trapped efficiently.
\end{itemize}

This study was largely possible due to the capabilities of the EPD instrument onboard \textit{SolO}, which observes in a wide range of energy channels and also provides directional measurements. Utilising multi-viewpoint remote sensing observations from widely distributed spacecraft and ground-based observations in many available wavelengths, it is possible to probe mechanisms continuously from deep in the solar corona to interplanetary space. Future studies must aim to include high resolution radio imaging from the LOw Frequency ARray \citep[LOFAR;][]{Haarlem13}, and the full capabilities of Solar Orbiter and Parker Solar Probe \citep[PSP;][]{Fox2016} to better understand the acceleration mechanisms at the Sun and the effects of coronal and interplanetary transport.


\begin{acknowledgements}
    This work received funding from the European Union's Horizon 2020 research and innovation program under grant agreement No. 870405 (EUHFORIA 2.0), and No. 101004159 (SERPENTINE). I.C.J.\ and J.M.\ acknowledge funding by the BRAIN-be project SWiM (Solar Wind Modelling with EUHFORIA for the new heliospheric missions). N.D.\ and I.C.J.\ are grateful for support by the Academy of Finland (SHOCKSEE, grant no.\ 346902). A.K.\ acknowledges financial support from NASA's NNN06AA01C (SO-SIS Phase-E) contract. Work in the University of Turku was performed under the umbrella of Finnish Centre of Excellence in Research of Sustainable Space (FORESAIL, Academy of Finland grant no.\ 336809). N.D.\ is grateful for support by the Turku Collegium for Science, Medicine and Technology of the University of Turku, Finland and support of Academy of Finland (SHOCKSEE, grant no.\ 346902).
    N.W.\ acknowledges support from the NASA program NNH17ZDA001N-LWS and from the Research Foundation -- Flanders (FWO -- Vlaanderen, fellowship no.\ 1184319N)
    T.W.\ acknowledges DLR grant 50 OC 2101. The work of F.S. was supported by DLR grant No. 50 OT 1904. I.C.J.\ thanks Dr. Vladimir Krasnoselskikh for the productive discussions regarding the various aspects of collisionless shocks.
    \textit{Solar Orbiter} is a space mission of international collaboration between ESA and NASA, operated by ESA. The STIX instrument is an international collaboration between Switzerland, Poland, France, Czech Republic, Germany, Austria, Ireland, and Italy. The authors also thank Dr. Milan Maksimovic for providing data products from the Radio Plasma Waves (RPW) instrument onboard \textit{Solar Orbiter}. EIT and LASCO data have  been used courtesy of the SOHO/EIT and SOHO/LASCO consortiums, respectively. The STEREO SECCHI data are produced by a consortium of RAL(UK), NRL(USA), LMSAL(USA), GSFC(USA), MPS(Germany), CSL(Belgium), IOTA(France), and IAS(France). The Wind/WAVES instrument was designed and built as a joint effort of the Paris-Meudon Observatory, the University of Minnesota, and the Goddard Space Flight Center, and the data are available at the instrument Web site. The authors are grateful to Dr. Xavier Bonnin for providing direction finding data from the WAVES experiment onboard Wind. We thank the radio monitoring service at LESIA (Observatoire de Paris) for providing value-added data that have been used for this study. HMI data are provided courtesy of NASA/SDO  science teams.
\end{acknowledgements}


\bibliographystyle{aa}
\bibliography{bibtex27092012}

\begin{thebibliography}{133}
\expandafter\ifx\csname natexlab\endcsname\relax\def\natexlab#1{#1}\fi

\bibitem[{{Agueda} {et~al.}(2014){Agueda}, {Klein}, {Vilmer},
  {Rodr{\'{\i}}guez-Gas{\'e}n}, {Malandraki}, {Papaioannou}, {Subir{\`a}},
  {Sanahuja}, {Valtonen}, {Dr{\"o}ge}, {Nindos}, {Heber}, {Braune}, {Usoskin},
  {Heynderickx}, {Talew}, \& {Vainio}}]{Agueda2014}
{Agueda}, N., {Klein}, K.-L., {Vilmer}, N., {et~al.} 2014, \aap, 570, A5

\bibitem[{{Anastasiadis} {et~al.}(2019){Anastasiadis}, {Lario}, {Papaioannou},
  {Kouloumvakos}, \& {Vourlidas}}]{Anastasiadis2019}
{Anastasiadis}, A., {Lario}, D., {Papaioannou}, A., {Kouloumvakos}, A., \&
  {Vourlidas}, A. 2019, Philosophical Transactions of the Royal Society of
  London Series A, 377, 20180100

\bibitem[{{Arge} {et~al.}(2010){Arge}, {Henney}, {Koller}, {Compeau}, {Young},
  {MacKenzie}, {Fay}, \& {Harvey}}]{Arge2009}
{Arge}, C.~N., {Henney}, C.~J., {Koller}, J., {et~al.} 2010, in Solar Wind 12,
  ed. M.~{Maksimovic}, K.~{Issautier}, N.~{Meyer-Vernet}, M.~M.~{Moncuquet}, \&
  F.~{Pantellini}, Vol. CS-1216, Am. Inst. Phys., Melville, 343--346

\bibitem[{{Aurass} {et~al.}(1998){Aurass}, {Hofmann}, \& {Urbarz}}]{Aurass98}
{Aurass}, H., {Hofmann}, A., \& {Urbarz}, H.~W. 1998, \aap, 334, 289

\bibitem[{{Badman} {et~al.}(2022){Badman}, {Carley}, {Ca{\~n}izares},
  {Dresing}, {Jian}, {Lario}, {Gallagher}, {Mart{\'\i}nez Oliveros}, {Pulupa},
  \& {Bale}}]{Badman2022}
{Badman}, S.~T., {Carley}, E., {Ca{\~n}izares}, L.~A., {et~al.} 2022, \apj,
  938, 95

\bibitem[{{Bale} {et~al.}(2016){Bale}, {Goetz}, {Harvey}, {Turin}, {Bonnell},
  {Dudok de Wit}, {Ergun}, {MacDowall}, {Pulupa}, {Andre}, {Bolton},
  {Bougeret}, {Bowen}, {Burgess}, {Cattell}, {Chandran}, {Chaston}, {Chen},
  {Choi}, {Connerney}, {Cranmer}, {Diaz-Aguado}, {Donakowski}, {Drake},
  {Farrell}, {Fergeau}, {Fermin}, {Fischer}, {Fox}, {Glaser}, {Goldstein},
  {Gordon}, {Hanson}, {Harris}, {Hayes}, {Hinze}, {Hollweg}, {Horbury},
  {Howard}, {Hoxie}, {Jannet}, {Karlsson}, {Kasper}, {Kellogg}, {Kien},
  {Klimchuk}, {Krasnoselskikh}, {Krucker}, {Lynch}, {Maksimovic}, {Malaspina},
  {Marker}, {Martin}, {Martinez-Oliveros}, {McCauley}, {McComas}, {McDonald},
  {Meyer-Vernet}, {Moncuquet}, {Monson}, {Mozer}, {Murphy}, {Odom},
  {Oliverson}, {Olson}, {Parker}, {Pankow}, {Phan}, {Quataert}, {Quinn},
  {Ruplin}, {Salem}, {Seitz}, {Sheppard}, {Siy}, {Stevens}, {Summers}, {Szabo},
  {Timofeeva}, {Vaivads}, {Velli}, {Yehle}, {Werthimer}, \& {Wygant}}]{Bale16}
{Bale}, S.~D., {Goetz}, K., {Harvey}, P.~R., {et~al.} 2016, \ssr, 204, 49

\bibitem[{{Bale} {et~al.}(1999){Bale}, {Reiner}, {Bougeret}, {Kaiser},
  {Krucker}, {Larson}, \& {Lin}}]{Bale99}
{Bale}, S.~D., {Reiner}, M.~J., {Bougeret}, J.~L., {et~al.} 1999, \grl, 26,
  1573

\bibitem[{{Balikhin} \& {Gedalin}(2022)}]{Balikhin22}
{Balikhin}, M. \& {Gedalin}, M. 2022, \apj, 925, 90

\bibitem[{{Ball} \& {Melrose}(2001)}]{Ball2001}
{Ball}, L. \& {Melrose}, D.~B. 2001, \pasa, 18, 361

\bibitem[{{Bastian} {et~al.}(1998){Bastian}, {Benz}, \& {Gary}}]{Bastian98}
{Bastian}, T.~S., {Benz}, A.~O., \& {Gary}, D.~E. 1998, \araa, 36, 131

\bibitem[{{Battaglia} {et~al.}(2021){Battaglia}, {Saqri}, {Massa},
  {Perracchione}, {Dickson}, {Xiao}, {Veronig}, {Warmuth}, {Battaglia},
  {Hurford}, {Meuris}, {Limousin}, {Etesi}, {Maloney}, {Schwartz}, {Kuhar},
  {Schuller}, {Senthamizh Pavai}, {Musset}, {Ryan}, {Kleint}, {Piana},
  {Massone}, {Benvenuto}, {Sylwester}, {Litwicka}, {St{\c{e}}{\'s}licki},
  {Mrozek}, {Vilmer}, {F{\'a}rn{\'\i}k}, {Ka{\v{s}}parov{\'a}}, {Mann},
  {Gallagher}, {Dennis}, {Csillaghy}, {Benz}, \& {Krucker}}]{Battaglia21}
{Battaglia}, A.~F., {Saqri}, J., {Massa}, P., {et~al.} 2021, \aap, 656, A4

\bibitem[{{Bell}(1978)}]{Bell78}
{Bell}, A.~R. 1978, \mnras, 182, 147

\bibitem[{{Benz} {et~al.}(2009){Benz}, {Monstein}, {Meyer}, {Manoharan},
  {Ramesh}, {Altyntsev}, {Lara}, {Paez}, \& {Cho}}]{Benz09}
{Benz}, A.~O., {Monstein}, C., {Meyer}, H., {et~al.} 2009, Earth Moon and
  Planets, 104, 277

\bibitem[{{Bougeret} {et~al.}(2008){Bougeret}, {Goetz}, {Kaiser}, {Bale},
  {Kellogg}, {Maksimovic}, {Monge}, {Monson}, {Astier}, {Davy}, {Dekkali},
  {Hinze}, {Manning}, {Aguilar-Rodriguez}, {Bonnin}, {Briand}, {Cairns},
  {Cattell}, {Cecconi}, {Eastwood}, {Ergun}, {Fainberg}, {Hoang}, {Huttunen},
  {Krucker}, {Lecacheux}, {MacDowall}, {Macher}, {Mangeney}, {Meetre},
  {Moussas}, {Nguyen}, {Oswald}, {Pulupa}, {Reiner}, {Robinson}, {Rucker},
  {Salem}, {Santolik}, {Silvis}, {Ullrich}, {Zarka}, \&
  {Zouganelis}}]{Bougeret08}
{Bougeret}, J.~L., {Goetz}, K., {Kaiser}, M.~L., {et~al.} 2008, \ssr, 136, 487

\bibitem[{{Bougeret} {et~al.}(1995){Bougeret}, {Kaiser}, {Kellogg}, {Manning},
  {Goetz}, {Monson}, {Monge}, {Friel}, {Meetre}, {Perche}, {Sitruk}, \&
  {Hoang}}]{Bougeret95}
{Bougeret}, J.-L., {Kaiser}, M.~L., {Kellogg}, P.~J., {et~al.} 1995, Space
  Science Reviews, 71, 231

\bibitem[{{Brown}(1971)}]{Brown71}
{Brown}, J.~C. 1971, \solphys, 18, 489

\bibitem[{{Brueckner} {et~al.}(1995){Brueckner}, {Howard}, {Koomen},
  {Korendyke}, {Michels}, {Moses}, {Socker}, {Dere}, {Lamy}, {Llebaria},
  {Bout}, {Schwenn}, {Simnett}, {Bedford}, \& {Eyles}}]{Brueckner95}
{Brueckner}, G.~E., {Howard}, R.~A., {Koomen}, M.~J., {et~al.} 1995, \solphys,
  162, 357

\bibitem[{{Cairns}(1987)}]{Cairns87}
{Cairns}, I.~H. 1987, Journal of Plasma Physics, 38, 169

\bibitem[{{Cairns} {et~al.}(2003){Cairns}, {Knock}, {Robinson}, \&
  {Kuncic}}]{Cairns03}
{Cairns}, I.~H., {Knock}, S.~A., {Robinson}, P.~A., \& {Kuncic}, Z. 2003, \ssr,
  107, 27

\bibitem[{{Cane} {et~al.}(1981){Cane}, {Stone}, {Fainberg}, {Steinberg},
  {Hoang}, \& {Stewart}}]{Cane81}
{Cane}, H.~V., {Stone}, R.~G., {Fainberg}, J., {et~al.} 1981, \grl, 8, 1285

\bibitem[{{Domingo} {et~al.}(1995){Domingo}, {Fleck}, \& {Poland}}]{Domingo95}
{Domingo}, V., {Fleck}, B., \& {Poland}, A.~I. 1995, \solphys, 162, 1

\bibitem[{{Dresing} {et~al.}(2020){Dresing}, {Effenberger},
  {G{\'o}mez-Herrero}, {Heber}, {Klassen}, {Kollhoff}, {Richardson}, \&
  {Theesen}}]{Dresing2020}
{Dresing}, N., {Effenberger}, F., {G{\'o}mez-Herrero}, R., {et~al.} 2020, \apj,
  889, 143

\bibitem[{{Dresing} {et~al.}(2022){Dresing}, {Kouloumvakos}, {Vainio}, \&
  {Rouillard}}]{Dresing22}
{Dresing}, N., {Kouloumvakos}, A., {Vainio}, R., \& {Rouillard}, A. 2022,
  \apjl, 925, L21

\bibitem[{{Dr{\"o}ge}(2000)}]{Droege2000}
{Dr{\"o}ge}, W. 2000, \apj, 537, 1073

\bibitem[{{Dulk} {et~al.}(2000){Dulk}, {Leblanc}, {Bastian}, \&
  {Bougeret}}]{Dulk2000}
{Dulk}, G.~A., {Leblanc}, Y., {Bastian}, T.~S., \& {Bougeret}, J.-L. 2000,
  \jgr, 105, 27343

\bibitem[{{F{\'a}rn{\'\i}k} {et~al.}(2003){F{\'a}rn{\'\i}k}, {Hudson},
  {Karlick{\'y}}, \& {Kosugi}}]{Farnik03}
{F{\'a}rn{\'\i}k}, F., {Hudson}, H.~S., {Karlick{\'y}}, M., \& {Kosugi}, T.
  2003, \aap, 399, 1159

\bibitem[{{Fox} {et~al.}(2016){Fox}, {Velli}, {Bale}, {Decker}, {Driesman},
  {Howard}, {Kasper}, {Kinnison}, {Kusterer}, {Lario}, {Lockwood}, {McComas},
  {Raouafi}, \& {Szabo}}]{Fox2016}
{Fox}, N.~J., {Velli}, M.~C., {Bale}, S.~D., {et~al.} 2016, \ssr, 204, 7

\bibitem[{{Galeev}(1984)}]{Galeev84}
{Galeev}, A.~A. 1984, Soviet Journal of Experimental and Theoretical Physics,
  59, 965

\bibitem[{{Galeev} {et~al.}(1995){Galeev}, {Malkov}, \& {V{\"o}lk}}]{Galeev95}
{Galeev}, A.~A., {Malkov}, M.~A., \& {V{\"o}lk}, H.~J. 1995, Journal of Plasma
  Physics, 54, 59

\bibitem[{{Garcia}(1994)}]{Howard94}
{Garcia}, H.~A. 1994, \solphys, 154, 275

\bibitem[{{Gedalin}(1996)}]{Gedalin96}
{Gedalin}, M. 1996, \jgr, 101, 4871

\bibitem[{Gieseler {et~al.}(2022)Gieseler, Dresing, Palmroos, {Freiherr von
  Forstner}, Price, Vainio, Kouloumvakos, Rodr\'iguez-Garc\'ia, Trotta,
  G\'enot, Masson, Roth, \& Veronig}]{Gieseler2022}
Gieseler, J., Dresing, N., Palmroos, C., {et~al.} 2022, Frontiers in Astronomy
  and Space Physics, (accepted)

\bibitem[{{Gieseler} {et~al.}(2000){Gieseler}, {Jones}, \& {Kang}}]{Gieseler00}
{Gieseler}, U.~D.~J., {Jones}, T.~W., \& {Kang}, H. 2000, \aap, 364, 911

\bibitem[{{Ginzburg} \& {Zhelezniakov}(1958)}]{Ginzburg58}
{Ginzburg}, V.~L. \& {Zhelezniakov}, V.~V. 1958, \sovast, 2, 653

\bibitem[{{Glesener} {et~al.}(2012){Glesener}, {Krucker}, \&
  {Lin}}]{Glesener2012}
{Glesener}, L., {Krucker}, S., \& {Lin}, R.~P. 2012, \apj, 754, 9

\bibitem[{{Grigis} \& {Benz}(2004)}]{Grigis04}
{Grigis}, P.~C. \& {Benz}, A.~O. 2004, \aap, 426, 1093

\bibitem[{{Grigis} \& {Benz}(2008)}]{Grigis08}
{Grigis}, P.~C. \& {Benz}, A.~O. 2008, \apj, 683, 1180

\bibitem[{{Harten} \& {Clark}(1995)}]{Harten1995}
{Harten}, R. \& {Clark}, K. 1995, \ssr, 71, 23

\bibitem[{{Howard} {et~al.}(2008){Howard}, {Moses}, {Vourlidas}, {Newmark},
  {Socker}, {Plunkett}, {Korendyke}, {Cook}, {Hurley}, {Davila}, {Thompson},
  {St Cyr}, {Mentzell}, {Mehalick}, {Lemen}, {Wuelser}, {Duncan}, {Tarbell},
  {Wolfson}, {Moore}, {Harrison}, {Waltham}, {Lang}, {Davis}, {Eyles},
  {Mapson-Menard}, {Simnett}, {Halain}, {Defise}, {Mazy}, {Rochus}, {Mercier},
  {Ravet}, {Delmotte}, {Auchere}, {Delaboudiniere}, {Bothmer}, {Deutsch},
  {Wang}, {Rich}, {Cooper}, {Stephens}, {Maahs}, {Baugh}, {McMullin}, \&
  {Carter}}]{Howard08}
{Howard}, R.~A., {Moses}, J.~D., {Vourlidas}, A., {et~al.} 2008, \ssr, 136, 67

\bibitem[{{Huttunen-Heikinmaa} {et~al.}(2005){Huttunen-Heikinmaa}, {Valtonen},
  \& {Laitinen}}]{Huttunen2005}
{Huttunen-Heikinmaa}, K., {Valtonen}, E., \& {Laitinen}, T. 2005, \aap, 442,
  673

\bibitem[{{Iwai} {et~al.}(2017){Iwai}, {Kubo}, {Ishibashi}, {Naoi}, {Harada},
  {Ema}, {Hayashi}, \& {Chikahiro}}]{Iwai17}
{Iwai}, K., {Kubo}, Y., {Ishibashi}, H., {et~al.} 2017, Earth, Planets and
  Space, 69, 95

\bibitem[{{Jebaraj} {et~al.}(2021){Jebaraj}, {Kouloumvakos}, {Magdalenic},
  {Rouillard}, {Mann}, {Krupar}, \& {Poedts}}]{Jebaraj21}
{Jebaraj}, I.~C., {Kouloumvakos}, A., {Magdalenic}, J., {et~al.} 2021, \aap,
  654, A64

\bibitem[{{Jebaraj} {et~al.}(2022){Jebaraj}, {Magdaleni{\'c}},
  {Krasnoselskikh}, {Krupar}, \& {Poedts}}]{Jebaraj22}
{Jebaraj}, I.~C., {Magdaleni{\'c}}, J., {Krasnoselskikh}, V., {Krupar}, V., \&
  {Poedts}, S. 2022, \aap, (in press)

\bibitem[{{Jebaraj} {et~al.}(2020){Jebaraj}, {Magdaleni{\'c}}, {Podladchikova},
  {Scolini}, {Pomoell}, {Veronig}, {Dissauer}, {Krupar}, {Kilpua}, \&
  {Poedts}}]{Jebaraj20}
{Jebaraj}, I.~C., {Magdaleni{\'c}}, J., {Podladchikova}, T., {et~al.} 2020,
  \aap, 639, A56

\bibitem[{{Kahler} \& {Hundhausen}(1992)}]{Kahler1992}
{Kahler}, S.~W. \& {Hundhausen}, A.~J. 1992, \jgr, 97, 1619

\bibitem[{{Kaiser}(2005)}]{Kaiser05}
{Kaiser}, M.~L. 2005, Advances in Space Research, 36, 1483

\bibitem[{{Kaiser} {et~al.}(2008){Kaiser}, {Kucera}, {Davila}, {St.~Cyr},
  {Guhathakurta}, \& {Christian}}]{Kaiser08}
{Kaiser}, M.~L., {Kucera}, T.~A., {Davila}, J.~M., {et~al.} 2008, \ssr, 136, 5

\bibitem[{{Kiplinger}(1995)}]{Kiplinger95}
{Kiplinger}, A.~L. 1995, \apj, 453, 973

\bibitem[{{Kirk} \& {Dendy}(2001)}]{Kirk01}
{Kirk}, J.~G. \& {Dendy}, R.~O. 2001, Journal of Physics G Nuclear Physics, 27,
  1589

\bibitem[{{Klassen} {et~al.}(2018){Klassen}, {Dresing}, {G{\'o}mez-Herrero},
  {Heber}, \& {Veronig}}]{Klassen18}
{Klassen}, A., {Dresing}, N., {G{\'o}mez-Herrero}, R., {Heber}, B., \&
  {Veronig}, A. 2018, \aap, 614, A61

\bibitem[{{Klassen} {et~al.}(2011{\natexlab{a}}){Klassen}, {G{\'o}mez-Herrero},
  \& {Heber}}]{Klassen11a}
{Klassen}, A., {G{\'o}mez-Herrero}, R., \& {Heber}, B. 2011{\natexlab{a}},
  \solphys, 273, 413

\bibitem[{{Klassen} {et~al.}(2012){Klassen}, {G{\'o}mez-Herrero}, {Heber},
  {Kartavykh}, {Dr{\"o}ge}, \& {Klein}}]{Klassen12}
{Klassen}, A., {G{\'o}mez-Herrero}, R., {Heber}, B., {et~al.} 2012, \aap, 542,
  A28

\bibitem[{{Klassen} {et~al.}(2011{\natexlab{b}}){Klassen}, {G{\'o}mez-Herrero},
  {M{\"u}ller-Mellin}, {Heber}, {Wimmer-Schweingruber}, {Opitz}, \&
  {Sauvaud}}]{Klassen11b}
{Klassen}, A., {G{\'o}mez-Herrero}, R., {M{\"u}ller-Mellin}, R., {et~al.}
  2011{\natexlab{b}}, \aap, 528, A84

\bibitem[{{Klein} {et~al.}(1999){Klein}, {Chupp}, {Trottet}, {Magun}, {Dunphy},
  {Rieger}, \& {Urpo}}]{Klein1999}
{Klein}, K.~L., {Chupp}, E.~L., {Trottet}, G., {et~al.} 1999, \aap, 348, 271

\bibitem[{{Klein} \& {Dalla}(2017)}]{Klein2017}
{Klein}, K.-L. \& {Dalla}, S. 2017, \ssr, 212, 1107

\bibitem[{{Klein} {et~al.}(2022){Klein}, {Musset}, {Vilmer}, {Briand},
  {Krucker}, {Francesco Battaglia}, {Dresing}, {Palmroos}, \& {Gary}}]{Klein22}
{Klein}, K.-L., {Musset}, S., {Vilmer}, N., {et~al.} 2022, \aap, 663, A173

\bibitem[{{Kong} {et~al.}(2016){Kong}, {Chen}, {Guo}, {Feng}, {Du}, \&
  {Li}}]{Kong16}
{Kong}, X., {Chen}, Y., {Guo}, F., {et~al.} 2016, \apj, 821, 32

\bibitem[{{Kong} {et~al.}(2017){Kong}, {Guo}, {Giacalone}, {Li}, \&
  {Chen}}]{Kong17}
{Kong}, X., {Guo}, F., {Giacalone}, J., {Li}, H., \& {Chen}, Y. 2017, \apj,
  851, 38

\bibitem[{{Kontar} \& {Reid}(2009)}]{Kontar09}
{Kontar}, E.~P. \& {Reid}, H. A.~S. 2009, \apjl, 695, L140

\bibitem[{{Kouloumvakos} {et~al.}(2022{\natexlab{a}}){Kouloumvakos}, {Kwon},
  {Rodr{\'\i}guez-Garc{\'\i}a}, {Lario}, {Dresing}, {Kilpua}, {Vainio},
  {T{\"o}r{\"o}k}, {Plotnikov}, {Rouillard}, {Downs}, {Linker}, {Malandraki},
  {Pinto}, {Riley}, \& {Allen}}]{Kouloumvakos2022A&A}
{Kouloumvakos}, A., {Kwon}, R.~Y., {Rodr{\'\i}guez-Garc{\'\i}a}, L., {et~al.}
  2022{\natexlab{a}}, \aap, 660, A84

\bibitem[{{Kouloumvakos} {et~al.}(2015){Kouloumvakos}, {Nindos}, {Valtonen},
  {Alissandrakis}, {Malandraki}, {Tsitsipis}, {Kontogeorgos}, {Moussas}, \&
  {Hillaris}}]{Kouloumvakos2015}
{Kouloumvakos}, A., {Nindos}, A., {Valtonen}, E., {et~al.} 2015, \aap, 580, A80

\bibitem[{{Kouloumvakos} {et~al.}(2022{\natexlab{b}}){Kouloumvakos},
  {Rodr{\'\i}guez-Garc{\'\i}a}, {Gieseler}, {Price}, {Vourlidas}, \&
  {Vainio}}]{Kouloumvakos2022}
{Kouloumvakos}, A., {Rodr{\'\i}guez-Garc{\'\i}a}, L., {Gieseler}, J., {et~al.}
  2022{\natexlab{b}}, Frontiers in Astronomy and Space Sciences, 9, 974137

\bibitem[{{Kouloumvakos} {et~al.}(2021){Kouloumvakos}, {Rouillard}, {Warmuth},
  {Magdaleni{\'c}}, {Jebaraj}, {Mann}, {Vainio}, \&
  {Monstein}}]{Kouloumvakos21}
{Kouloumvakos}, A., {Rouillard}, A., {Warmuth}, A., {et~al.} 2021, \apj, 913,
  99

\bibitem[{{Kouloumvakos} {et~al.}(2019){Kouloumvakos}, {Rouillard}, {Wu},
  {Vainio}, {Vourlidas}, {Plotnikov}, {Afanasiev}, \&
  {{\"O}nel}}]{Kouloumvakos19}
{Kouloumvakos}, A., {Rouillard}, A.~P., {Wu}, Y., {et~al.} 2019, \apj, 876, 80

\bibitem[{{Krasnoselskikh} {et~al.}(2019){Krasnoselskikh}, {Voshchepynets}, \&
  {Maksimovic}}]{Krasnoselskikh19}
{Krasnoselskikh}, V., {Voshchepynets}, A., \& {Maksimovic}, M. 2019, \apj, 879,
  51

\bibitem[{{Krasnoselskikh} {et~al.}(1991){Krasnoselskikh}, {Balikhin},
  {Alleyne}, {Klimov}, {Mier-Jedrzejowicz}, {Pardaens}, {Petrukovich},
  {Southwood}, {Vinogradova}, \& {Woolliscroft}}]{Krasnoselskikh91}
{Krasnoselskikh}, V.~V., {Balikhin}, M.~A., {Alleyne}, H. S.~C., {et~al.} 1991,
  Advances in Space Research, 11, 15

\bibitem[{{Krasnoselskikh} {et~al.}(1985){Krasnoselskikh}, {Kruchina},
  {Volokitin}, \& {Thejappa}}]{Krasnoselskikh85}
{Krasnoselskikh}, V.~V., {Kruchina}, E.~N., {Volokitin}, A.~S., \& {Thejappa},
  G. 1985, \aap, 149, 323

\bibitem[{{Krauss-Varban} \& {Wu}(1989)}]{KraussV1989a}
{Krauss-Varban}, D. \& {Wu}, C.~S. 1989, \jgr, 94, 15367

\bibitem[{{Krucker} {et~al.}(2020){Krucker}, {Hurford}, {Grimm}, {K{\"o}gl},
  {Gr{\"o}belbauer}, {Etesi}, {Casadei}, {Csillaghy}, {Benz}, {Arnold},
  {Molendini}, {Orleanski}, {Schori}, {Xiao}, {Kuhar}, {Hochmuth}, {Felix},
  {Schramka}, {Marcin}, {Kobler}, {Iseli}, {Dreier}, {Wiehl}, {Kleint},
  {Battaglia}, {Lastufka}, {Sathiapal}, {Lapadula}, {Bednarzik}, {Birrer},
  {Stutz}, {Wild}, {Marone}, {Skup}, {Cichocki}, {Ber}, {Rutkowski}, {Bujwan},
  {Juchnikowski}, {Winkler}, {Darmetko}, {Michalska}, {Seweryn}, {Bia{\l}ek},
  {Osica}, {Sylwester}, {Kowalinski}, {{\'S}cis{\l}owski}, {Siarkowski},
  {St{\k{e}}{\'s}licki}, {Mrozek}, {Podg{\'o}rski}, {Meuris}, {Limousin},
  {Gevin}, {Le Mer}, {Brun}, {Strugarek}, {Vilmer}, {Musset}, {Maksimovi{\'c}},
  {F{\'a}rn{\'\i}k}, {Koz{\'a}{\v{c}}ek}, {Ka{\v{s}}parov{\'a}}, {Mann},
  {{\"O}nel}, {Warmuth}, {Rendtel}, {Anderson}, {Bauer}, {Dionies}, {Paschke},
  {Pl{\"u}schke}, {Woche}, {Schuller}, {Veronig}, {Dickson}, {Gallagher},
  {Maloney}, {Bloomfield}, {Piana}, {Massone}, {Benvenuto}, {Massa},
  {Schwartz}, {Dennis}, {van Beek}, {Rodr{\'\i}guez-Pacheco}, \&
  {Lin}}]{Krucker20}
{Krucker}, S., {Hurford}, G.~J., {Grimm}, O., {et~al.} 2020, \aap, 642, A15

\bibitem[{{Krucker} {et~al.}(2011){Krucker}, {Kontar}, {Christe}, {Glesener},
  \& {Lin}}]{Krucker2011}
{Krucker}, S., {Kontar}, E.~P., {Christe}, S., {Glesener}, L., \& {Lin}, R.~P.
  2011, \apj, 742, 82

\bibitem[{{Krucker} {et~al.}(2007){Krucker}, {Kontar}, {Christe}, \&
  {Lin}}]{Krucker07}
{Krucker}, S., {Kontar}, E.~P., {Christe}, S., \& {Lin}, R.~P. 2007, \apjl,
  663, L109

\bibitem[{{Krucker} {et~al.}(2008){Krucker}, {Saint-Hilaire}, {Christe},
  {White}, {Chavier}, {Bale}, \& {Lin}}]{Krucker08}
{Krucker}, S., {Saint-Hilaire}, P., {Christe}, S., {et~al.} 2008, \apj, 681,
  644

\bibitem[{{Krupar} {et~al.}(2016){Krupar}, {Eastwood}, {Kruparova}, {Santolik},
  {Soucek}, {Magdaleni{\'c}}, {Vourlidas}, {Maksimovic}, {Bonnin}, {Bothmer},
  {Mrotzek}, {Pluta}, {Barnes}, {Davies}, {Mart{\'{\i}}nez Oliveros}, \&
  {Bale}}]{Krupar16}
{Krupar}, V., {Eastwood}, J.~P., {Kruparova}, O., {et~al.} 2016, \apjl, 823

\bibitem[{{Krupar} {et~al.}(2014){Krupar}, {Maksimovic}, {Santolik}, {Kontar},
  {Cecconi}, {Hoang}, {Kruparova}, {Soucek}, {Reid}, \&
  {Zaslavsky}}]{Krupar14a}
{Krupar}, V., {Maksimovic}, M., {Santolik}, O., {et~al.} 2014, \solphys, 289,
  3121

\bibitem[{{Krupar} {et~al.}(2020){Krupar}, {Szabo}, {Maksimovic}, {Kruparova},
  {Kontar}, {Balmaceda}, {Bonnin}, {Bale}, {Pulupa}, {Malaspina}, {Bonnell},
  {Harvey}, {Goetz}, {Dudok de Wit}, {MacDowall}, {Kasper}, {Case}, {Korreck},
  {Larson}, {Livi}, {Stevens}, {Whittlesey}, \& {Hegedus}}]{Krupar20}
{Krupar}, V., {Szabo}, A., {Maksimovic}, M., {et~al.} 2020, \apjs, 246, 57

\bibitem[{{Lario} {et~al.}(2022){Lario}, {Wijsen}, {Kwon}, {S{\'a}nchez-Cano},
  {Richardson}, {Pacheco}, {Palmerio}, {Stevens}, {Szabo}, {Heyner}, {Dresing},
  {G{\'o}mez-Herrero}, {Carcaboso}, {Aran}, {Afanasiev}, {Vainio}, {Riihonen},
  {Poedts}, {Br{\"u}den}, {Xu}, \& {Kollhoff}}]{Lario22}
{Lario}, D., {Wijsen}, N., {Kwon}, R.~Y., {et~al.} 2022, \apj, 934, 55

\bibitem[{{Lemen} {et~al.}(2012){Lemen}, {Title}, {Akin}, {Boerner}, {Chou},
  {Drake}, {Duncan}, {Edwards}, {Friedlaender}, {Heyman}, {Hurlburt}, {Katz},
  {Kushner}, {Levay}, {Lindgren}, {Mathur}, {McFeaters}, {Mitchell}, {Rehse},
  {Schrijver}, {Springer}, {Stern}, {Tarbell}, {Wuelser}, {Wolfson}, {Yanari},
  {Bookbinder}, {Cheimets}, {Caldwell}, {Deluca}, {Gates}, {Golub}, {Park},
  {Podgorski}, {Bush}, {Scherrer}, {Gummin}, {Smith}, {Auker}, {Jerram},
  {Pool}, {Soufli}, {Windt}, {Beardsley}, {Clapp}, {Lang}, \&
  {Waltham}}]{Lemen12}
{Lemen}, J.~R., {Title}, A.~M., {Akin}, D.~J., {et~al.} 2012, \solphys, 275, 17

\bibitem[{{Leroy} \& {Mangeney}(1984)}]{Leroy84}
{Leroy}, M.~M. \& {Mangeney}, A. 1984, Annales Geophysicae, 2, 449

\bibitem[{{Lin} {et~al.}(1995){Lin}, {Anderson}, {Ashford}, {Carlson},
  {Curtis}, {Ergun}, {Larson}, {McFadden}, {McCarthy}, {Parks}, {R{\`e}me},
  {Bosqued}, {Coutelier}, {Cotin}, {D'Uston}, {Wenzel}, {Sanderson}, {Henrion},
  {Ronnet}, \& {Paschmann}}]{Lin1995}
{Lin}, R.~P., {Anderson}, K.~A., {Ashford}, S., {et~al.} 1995, \ssr, 71, 125

\bibitem[{{Long} {et~al.}(2017){Long}, {Bloomfield}, {Chen}, {Downs},
  {Gallagher}, {Kwon}, {Vanninathan}, {Veronig}, {Vourlidas}, {Vr{\v s}nak},
  {Warmuth}, \& {{\v Z}ic}}]{Long17}
{Long}, D.~M., {Bloomfield}, D.~S., {Chen}, P.~F., {et~al.} 2017, \solphys,
  292, 7

\bibitem[{Lucas(1985)}]{lucas85}
Lucas, J.~M. 1985, Technometrics, 27, 129

\bibitem[{{Magdaleni{\'c}} {et~al.}(2014){Magdaleni{\'c}}, {Marqu{\'e}},
  {Krupar}, {Mierla}, {Zhukov}, {Rodriguez}, {Maksimovi{\'c}}, \&
  {Cecconi}}]{Magdalenic14}
{Magdaleni{\'c}}, J., {Marqu{\'e}}, C., {Krupar}, V., {et~al.} 2014, \apj, 791,
  115

\bibitem[{{Maksimovic} {et~al.}(2020){Maksimovic}, {Bale}, {Chust},
  {Khotyaintsev}, {Krasnoselskikh}, {Kretzschmar}, {Plettemeier}, {Rucker},
  {Sou{\v{c}}ek}, {Steller}, {{\v{S}}tver{\'a}k}, {Tr{\'a}vn{\'\i}{\v{c}}ek},
  {Vaivads}, {Chaintreuil}, {Dekkali}, {Alexandrova}, {Astier}, {Barbary},
  {B{\'e}rard}, {Bonnin}, {Boughedada}, {Cecconi}, {Chapron}, {Chariet},
  {Collin}, {de Conchy}, {Dias}, {Gu{\'e}guen}, {Lamy}, {Leray}, {Lion},
  {Malac-Allain}, {Matteini}, {Nguyen}, {Pantellini}, {Parisot}, {Plasson},
  {Thijs}, {Vecchio}, {Fratter}, {Bellouard}, {Lorf{\`e}vre}, {Danto},
  {Julien}, {Guilhem}, {Fiachetti}, {Sanisidro}, {Laffaye}, {Gonzalez},
  {Pontet}, {Qu{\'e}ruel}, {Jannet}, {Fergeau}, {Brochot}, {Cassam-Chenai},
  {Dudok de Wit}, {Timofeeva}, {Vincent}, {Agrapart}, {Delory}, {Turin},
  {Jeandet}, {Leroy}, {Pellion}, {Bouzid}, {Katra}, {Piberne}, {Recart},
  {Santol{\'\i}k}, {Kolma{\v{s}}ov{\'a}}, {Krupa{\v{r}}},
  {Krupa{\v{r}}ov{\'a}}, {P{\'\i}{\v{s}}a}, {Uhl{\'\i}{\v{r}}}, {L{\'a}n},
  {Ba{\v{s}}e}, {Ahl{\`e}n}, {Andr{\'e}}, {Bylander}, {Cripps}, {Cully},
  {Eriksson}, {Jansson}, {Johansson}, {Karlsson}, {Puccio},
  {B{\v{r}}{\'\i}nek}, {{\"O}ttacher}, {Panchenko}, {Berthomier}, {Goetz},
  {Hellinger}, {Horbury}, {Issautier}, {Kontar}, {Krucker}, {Le Contel},
  {Louarn}, {Martinovi{\'c}}, {Owen}, {Retino}, {Rodr{\'\i}guez-Pacheco},
  {Sahraoui}, {Wimmer-Schweingruber}, {Zaslavsky}, \&
  {Zouganelis}}]{Maksimovic20b}
{Maksimovic}, M., {Bale}, S.~D., {Chust}, T., {et~al.} 2020, \aap, 642, A12

\bibitem[{{Maksimovic} {et~al.}(1997){Maksimovic}, {Pierrard}, \&
  {Lemaire}}]{Maksimovic97}
{Maksimovic}, M., {Pierrard}, V., \& {Lemaire}, J.~F. 1997, \aap, 324, 725

\bibitem[{{Mann} \& {Klassen}(2005)}]{Mann05}
{Mann}, G. \& {Klassen}, A. 2005, \aap, 441, 319

\bibitem[{{Mann} {et~al.}(2009){Mann}, {Warmuth}, \& {Aurass}}]{Mann09}
{Mann}, G., {Warmuth}, A., \& {Aurass}, H. 2009, \aap, 494, 669

\bibitem[{{Massa} {et~al.}(2022){Massa}, {Battaglia}, {Volpara}, {Collier},
  {Hurford}, {Kuhar}, {Perracchione}, {Garbarino}, {Massone}, {Benvenuto},
  {Schuller}, {Warmuth}, {Dickson}, {Xiao}, {Maloney}, {Ryan}, {Piana}, \&
  {Krucker}}]{Massa22}
{Massa}, P., {Battaglia}, A.~F., {Volpara}, A., {et~al.} 2022, \solphys, 297,
  93

\bibitem[{{Massa} {et~al.}(2020){Massa}, {Schwartz}, {Tolbert}, {Massone},
  {Dennis}, {Piana}, \& {Benvenuto}}]{Massa20}
{Massa}, P., {Schwartz}, R., {Tolbert}, A.~K., {et~al.} 2020, \apj, 894, 46

\bibitem[{{Masson} {et~al.}(2019){Masson}, {Antiochos}, \&
  {DeVore}}]{Masson2019}
{Masson}, S., {Antiochos}, S.~K., \& {DeVore}, C.~R. 2019, \apj, 884, 143

\bibitem[{{McClements} {et~al.}(1997){McClements}, {Dendy}, {Bingham}, {Kirk},
  \& {Drury}}]{McClements97}
{McClements}, K.~G., {Dendy}, R.~O., {Bingham}, R., {Kirk}, J.~G., \& {Drury},
  L.~O. 1997, \mnras, 291, 241

\bibitem[{{McLean} \& {Labrum}(1985)}]{Mclean85}
{McLean}, D.~J. \& {Labrum}, N.~R. 1985, {Solar radiophysics : studies of
  emission from the sun at metre wavelengths}

\bibitem[{{Melrose}(2017)}]{Melrose17}
{Melrose}, D.~B. 2017, Reviews of Modern Plasma Physics, 1, 5

\bibitem[{{M{\"u}ller} {et~al.}(2020){M{\"u}ller}, {St. Cyr}, {Zouganelis},
  {Gilbert}, {Marsden}, {Nieves-Chinchilla}, {Antonucci}, {Auch{\`e}re},
  {Berghmans}, {Horbury}, {Howard}, {Krucker}, {Maksimovic}, {Owen}, {Rochus},
  {Rodriguez-Pacheco}, {Romoli}, {Solanki}, {Bruno}, {Carlsson}, {Fludra},
  {Harra}, {Hassler}, {Livi}, {Louarn}, {Peter}, {Sch{\"u}hle}, {Teriaca}, {del
  Toro Iniesta}, {Wimmer-Schweingruber}, {Marsch}, {Velli}, {De Groof},
  {Walsh}, \& {Williams}}]{Mueller20}
{M{\"u}ller}, D., {St. Cyr}, O.~C., {Zouganelis}, I., {et~al.} 2020, \aap, 642,
  A1

\bibitem[{{M{\"u}ller-Mellin} {et~al.}(2008){M{\"u}ller-Mellin},
  {B{\"o}ttcher}, {Falenski}, {Rode}, {Duvet}, {Sanderson}, {Butler},
  {Johlander}, \& {Smit}}]{Mueller2008}
{M{\"u}ller-Mellin}, R., {B{\"o}ttcher}, S., {Falenski}, J., {et~al.} 2008,
  \ssr, 136, 363

\bibitem[{{Musset} {et~al.}(2020){Musset}, {Jeunon}, \&
  {Glesener}}]{Musset2020}
{Musset}, S., {Jeunon}, M., \& {Glesener}, L. 2020, \apj, 889, 183

\bibitem[{{Newkirk}(1961)}]{Newkirk61}
{Newkirk}, G.~J. 1961, The Astrophysical Journal, 133, 983

\bibitem[{{Nindos}(2020)}]{Nindos20}
{Nindos}, A. 2020, Frontiers in Astronomy and Space Sciences, 7, 57

\bibitem[{Page(1954)}]{page54}
Page, E.~S. 1954, Biometrika, 41, 100

\bibitem[{{Papaioannou} {et~al.}(2016){Papaioannou}, {Sandberg},
  {Anastasiadis}, {Kouloumvakos}, {Georgoulis}, {Tziotziou}, {Tsiropoula},
  {Jiggens}, \& {Hilgers}}]{Papaioannou16}
{Papaioannou}, A., {Sandberg}, I., {Anastasiadis}, A., {et~al.} 2016, Journal
  of Space Weather and Space Climate, 6, A42

\bibitem[{{Pesnell} {et~al.}(2012){Pesnell}, {Thompson}, \&
  {Chamberlin}}]{Pesnell12}
{Pesnell}, W.~D., {Thompson}, B.~J., \& {Chamberlin}, P.~C. 2012, \solphys,
  275, 3

\bibitem[{{Petrosian}(2016)}]{Petrosian16}
{Petrosian}, V. 2016, \apj, 830, 28

\bibitem[{{Pohjolainen}(2008)}]{Pohjolainen08b}
{Pohjolainen}, S. 2008, \aap, 483, 297

\bibitem[{{Pulupa} {et~al.}(2017){Pulupa}, {Bale}, {Bonnell}, {Bowen},
  {Carruth}, {Goetz}, {Gordon}, {Harvey}, {Maksimovic},
  {Mart{\'\i}nez-Oliveros}, {Moncuquet}, {Saint-Hilaire}, {Seitz}, \&
  {Sundkvist}}]{Pulupa17}
{Pulupa}, M., {Bale}, S.~D., {Bonnell}, J.~W., {et~al.} 2017, Journal of
  Geophysical Research (Space Physics), 122, 2836

\bibitem[{{Ramesh} {et~al.}(2022){Ramesh}, {Kathiravan}, \&
  {Chellasamy}}]{Ramesh22}
{Ramesh}, R., {Kathiravan}, C., \& {Chellasamy}, E.~E. 2022, \apj, 932, 48

\bibitem[{{Reames}(2021)}]{Reames2021}
{Reames}, D.~V. 2021, {Solar Energetic Particles. A Modern Primer on
  Understanding Sources, Acceleration and Propagation}, Vol. 978

\bibitem[{{Reid} \& {Ratcliffe}(2014)}]{Reid14Review}
{Reid}, H. A.~S. \& {Ratcliffe}, H. 2014, Research in Astronomy and
  Astrophysics, 14, 773

\bibitem[{{Rodr{\'\i}guez-Pacheco} {et~al.}(2020){Rodr{\'\i}guez-Pacheco},
  {Wimmer-Schweingruber}, {Mason}, {Ho}, {S{\'a}nchez-Prieto}, {Prieto},
  {Mart{\'\i}n}, {Seifert}, {Andrews}, {Kulkarni}, {Panitzsch}, {Boden},
  {B{\"o}ttcher}, {Cernuda}, {Elftmann}, {Espinosa Lara}, {G{\'o}mez-Herrero},
  {Terasa}, {Almena}, {Begley}, {B{\"o}hm}, {Blanco}, {Boogaerts}, {Carrasco},
  {Castillo}, {da Silva Fari{\~n}a}, {de Manuel Gonz{\'a}lez}, {Drews},
  {Dupont}, {Eldrum}, {Gordillo}, {Guti{\'e}rrez}, {Haggerty}, {Hayes},
  {Heber}, {Hill}, {J{\"u}ngling}, {Kerem}, {Knierim}, {K{\"o}hler}, {Kolbe},
  {Kulemzin}, {Lario}, {Lees}, {Liang}, {Mart{\'\i}nez Hell{\'\i}n}, {Meziat},
  {Montalvo}, {Nelson}, {Parra}, {Paspirgilis}, {Ravanbakhsh}, {Richards},
  {Rodr{\'\i}guez-Polo}, {Russu}, {S{\'a}nchez}, {Schlemm}, {Schuster},
  {Seimetz}, {Steinhagen}, {Tammen}, {Tyagi}, {Varela}, {Yedla}, {Yu},
  {Agueda}, {Aran}, {Horbury}, {Klecker}, {Klein}, {Kontar}, {Krucker},
  {Maksimovic}, {Malandraki}, {Owen}, {Pacheco}, {Sanahuja}, {Vainio},
  {Connell}, {Dalla}, {Dr{\"o}ge}, {Gevin}, {Gopalswamy}, {Kartavykh},
  {Kudela}, {Limousin}, {Makela}, {Mann}, {{\"O}nel}, {Posner}, {Ryan},
  {Soucek}, {Hofmeister}, {Vilmer}, {Walsh}, {Wang}, {Wiedenbeck}, {Wirth}, \&
  {Zong}}]{Pacheco2020}
{Rodr{\'\i}guez-Pacheco}, J., {Wimmer-Schweingruber}, R.~F., {Mason}, G.~M.,
  {et~al.} 2020, \aap, 642, A7

\bibitem[{{Saito}(1970)}]{Saito70}
{Saito}, K. 1970, Annals of the Tokyo Astronomical Observatory, 12, 51

\bibitem[{{Schmidt} \& {Cairns}(2016)}]{Schmidt2016}
{Schmidt}, J.~M. \& {Cairns}, I.~H. 2016, \grl, 43, 50

\bibitem[{{Schrijver} \& {De Rosa}(2003)}]{Schrijver03}
{Schrijver}, C.~J. \& {De Rosa}, M.~L. 2003, \solphys, 212, 165

\bibitem[{{Stewart} \& {Magun}(1980)}]{Stewart80}
{Stewart}, R.~T. \& {Magun}, A. 1980, \pasa, 4, 53

\bibitem[{{Strauss} {et~al.}(2020){Strauss}, {Dresing}, {Kollhoff}, \&
  {Br{\"u}dern}}]{Strauss2020}
{Strauss}, R.~D., {Dresing}, N., {Kollhoff}, A., \& {Br{\"u}dern}, M. 2020,
  \apj, 897, 24

\bibitem[{{Thejappa}(1987)}]{Thejappa87}
{Thejappa}, G. 1987, \solphys, 111, 45

\bibitem[{{Tkachenko} {et~al.}(2021){Tkachenko}, {Krasnoselskikh}, \&
  {Voshchepynets}}]{Tkachenko21}
{Tkachenko}, A., {Krasnoselskikh}, V., \& {Voshchepynets}, A. 2021, \apj, 908,
  126

\bibitem[{{Treumann} \& {Jaroschek}(2008)}]{Treumann08para}
{Treumann}, R.~A. \& {Jaroschek}, C.~H. 2008, arXiv e-prints, arXiv:0805.2579

\bibitem[{{Tsytovich}(1966)}]{Tsytovich66}
{Tsytovich}, V.~N. 1966, Soviet Physics Uspekhi, 9, 370

\bibitem[{{Tsytovich}(1973)}]{Tsytovich73}
{Tsytovich}, V.~N. 1973, \araa, 11, 363

\bibitem[{{Uchida} {et~al.}(1973){Uchida}, {Altschuler}, \&
  {Newkirk}}]{Uchida73}
{Uchida}, Y., {Altschuler}, M.~D., \& {Newkirk}, Gordon, J. 1973, \solphys, 28,
  495

\bibitem[{{Vainio} {et~al.}(2013){Vainio}, {Valtonen}, {Heber}, {Malandraki},
  {Papaioannou}, {Klein}, {Afanasiev}, {Agueda}, {Aurass}, {Battarbee},
  {Braune}, {Dr{\"o}ge}, {Ganse}, {Hamadache}, {Heynderickx},
  {Huttunen-Heikinmaa}, {Kiener}, {Kilian}, {Kopp}, {Kouloumvakos}, {Maisala},
  {Mishev}, {Miteva}, {Nindos}, {Oittinen}, {Raukunen}, {Riihonen},
  {Rodr{\'\i}guez-Gas{\'e}n}, {Saloniemi}, {Sanahuja}, {Scherer}, {Spanier},
  {Tatischeff}, {Tziotziou}, {Usoskin}, \& {Vilmer}}]{Vainio13}
{Vainio}, R., {Valtonen}, E., {Heber}, B., {et~al.} 2013, Journal of Space
  Weather and Space Climate, 3, A12

\bibitem[{{Vaisberg} {et~al.}(1983){Vaisberg}, {Galeev}, {Zastenker}, {Klimov},
  {Nozdrachev}, {Sagdeev}, {Sokolov}, \& {Shapiro}}]{Vaisberg83}
{Vaisberg}, D.~L., {Galeev}, A.~A., {Zastenker}, G.~N., {et~al.} 1983, Soviet
  Journal of Experimental and Theoretical Physics, 58, 716

\bibitem[{{van Haarlem} {et~al.}(2013){van Haarlem}, {Wise}, {Gunst}, {Heald},
  {McKean}, {Hessels}, {de Bruyn}, {Nijboer}, {Swinbank}, {Fallows},
  {Brentjens}, {Nelles}, {Beck}, {Falcke}, {Fender}, {H{\"o}randel},
  {Koopmans}, {Mann}, {Miley}, {R{\"o}ttgering}, {Stappers}, {Wijers},
  {Zaroubi}, {van den Akker}, {Alexov}, {Anderson}, {Anderson}, {van Ardenne},
  {Arts}, {Asgekar}, {Avruch}, {Batejat}, {B{\"a}hren}, {Bell}, {Bell}, {van
  Bemmel}, {Bennema}, {Bentum}, {Bernardi}, {Best}, {B{\^\i}rzan}, {Bonafede},
  {Boonstra}, {Braun}, {Bregman}, {Breitling}, {van de Brink}, {Broderick},
  {Broekema}, {Brouw}, {Br{\"u}ggen}, {Butcher}, {van Cappellen}, {Ciardi},
  {Coenen}, {Conway}, {Coolen}, {Corstanje}, {Damstra}, {Davies}, {Deller},
  {Dettmar}, {van Diepen}, {Dijkstra}, {Donker}, {Doorduin}, {Dromer}, {Drost},
  {van Duin}, {Eisl{\"o}ffel}, {van Enst}, {Ferrari}, {Frieswijk}, {Gankema},
  {Garrett}, {de Gasperin}, {Gerbers}, {de Geus}, {Grie{\ss}meier}, {Grit},
  {Gruppen}, {Hamaker}, {Hassall}, {Hoeft}, {Holties}, {Horneffer}, {van der
  Horst}, {van Houwelingen}, {Huijgen}, {Iacobelli}, {Intema}, {Jackson},
  {Jelic}, {de Jong}, {Juette}, {Kant}, {Karastergiou}, {Koers}, {Kollen},
  {Kondratiev}, {Kooistra}, {Koopman}, {Koster}, {Kuniyoshi}, {Kramer},
  {Kuper}, {Lambropoulos}, {Law}, {van Leeuwen}, {Lemaitre}, {Loose}, {Maat},
  {Macario}, {Markoff}, {Masters}, {McFadden}, {McKay-Bukowski}, {Meijering},
  {Meulman}, {Mevius}, {Middelberg}, {Millenaar}, {Miller-Jones}, {Mohan},
  {Mol}, {Morawietz}, {Morganti}, {Mulcahy}, {Mulder}, {Munk}, {Nieuwenhuis},
  {van Nieuwpoort}, {Noordam}, {Norden}, {Noutsos}, {Offringa}, {Olofsson},
  {Omar}, {Orr{\'u}}, {Overeem}, {Paas}, {Pand ey-Pommier}, {Pandey}, {Pizzo},
  {Polatidis}, {Rafferty}, {Rawlings}, {Reich}, {de Reijer}, {Reitsma},
  {Renting}, {Riemers}, {Rol}, {Romein}, {Roosjen}, {Ruiter}, {Scaife}, {van
  der Schaaf}, {Scheers}, {Schellart}, {Schoenmakers}, {Schoonderbeek},
  {Serylak}, {Shulevski}, {Sluman}, {Smirnov}, {Sobey}, {Spreeuw}, {Steinmetz},
  {Sterks}, {Stiepel}, {Stuurwold}, {Tagger}, {Tang}, {Tasse}, {Thomas},
  {Thoudam}, {Toribio}, {van der Tol}, {Usov}, {van Veelen}, {van der Veen},
  {ter Veen}, {Verbiest}, {Vermeulen}, {Vermaas}, {Vocks}, {Vogt}, {de Vos},
  {van der Wal}, {van Weeren}, {Weggemans}, {Weltevrede}, {White}, {Wijnholds},
  {Wilhelmsson}, {Wucknitz}, {Yatawatta}, {Zarka}, {Zensus}, \& {van
  Zwieten}}]{Haarlem13}
{van Haarlem}, M.~P., {Wise}, M.~W., {Gunst}, A.~W., {et~al.} 2013, \aap, 556,
  A2

\bibitem[{{Vlahos} {et~al.}(2019){Vlahos}, {Anastasiadis}, {Papaioannou},
  {Kouloumvakos}, \& {Isliker}}]{Vlahos2019}
{Vlahos}, L., {Anastasiadis}, A., {Papaioannou}, A., {Kouloumvakos}, A., \&
  {Isliker}, H. 2019, Philosophical Transactions of the Royal Society of London
  Series A, 377, 20180095

\bibitem[{{von Rosenvinge} {et~al.}(2008){von Rosenvinge}, {Reames}, {Baker},
  {Hawk}, {Nolan}, {Ryan}, {Shuman}, {Wortman}, {Mewaldt}, {Cummings}, {Cook},
  {Labrador}, {Leske}, \& {Wiedenbeck}}]{Rosenvinge2008}
{von Rosenvinge}, T.~T., {Reames}, D.~V., {Baker}, R., {et~al.} 2008, \ssr,
  136, 391

\bibitem[{{Vr{\v{s}}nak} \& {Luli{\'c}}(2000{\natexlab{a}})}]{Vrsnak00a}
{Vr{\v{s}}nak}, B. \& {Luli{\'c}}, S. 2000{\natexlab{a}}, \solphys, 196, 157

\bibitem[{{Vr{\v{s}}nak} \& {Luli{\'c}}(2000{\natexlab{b}})}]{Vrsnak00b}
{Vr{\v{s}}nak}, B. \& {Luli{\'c}}, S. 2000{\natexlab{b}}, \solphys, 196, 181

\bibitem[{{Warmuth}(2015)}]{Warmuth15}
{Warmuth}, A. 2015, Living Reviews in Solar Physics, 12, 3

\bibitem[{{Warmuth} \& {Mann}(2005)}]{Warmuth05}
{Warmuth}, A. \& {Mann}, G. 2005, \aap, 435, 1123

\bibitem[{{Warmuth} \& {Mann}(2020)}]{Warmuth20a}
{Warmuth}, A. \& {Mann}, G. 2020, \aap, 644, A172

\bibitem[{{Warmuth} {et~al.}(2009){Warmuth}, {Mann}, \& {Aurass}}]{Warmuth09b}
{Warmuth}, A., {Mann}, G., \& {Aurass}, H. 2009, \aap, 494, 677

\bibitem[{{Warmuth} {et~al.}(2020){Warmuth}, {{\"O}nel}, {Mann}, {Rendtel},
  {Strassmeier}, {Denker}, {Hurford}, {Krucker}, {Anderson}, {Bauer},
  {Bittner}, {Dionies}, {Paschke}, {Pl{\"u}schke}, {Sablowski}, {Schuller},
  {Senthamizh Pavai}, {Woche}, {Casadei}, {K{\"o}gl}, {Arnold},
  {Gr{\"o}belbauer}, {Schori}, {Wiehl}, {Csillaghy}, {Grimm}, {Orleanski},
  {Skup}, {Bujwan}, {Rutkowski}, \& {Ber}}]{Warmuth20b}
{Warmuth}, A., {{\"O}nel}, H., {Mann}, G., {et~al.} 2020, \solphys, 295, 90

\bibitem[{{Wiegelmann} {et~al.}(2006){Wiegelmann}, {Inhester}, \&
  {Sakurai}}]{2006SoPh..233..215W}
{Wiegelmann}, T., {Inhester}, B., \& {Sakurai}, T. 2006, \solphys, 233, 215

\bibitem[{{Wiegelmann} {et~al.}(2012){Wiegelmann}, {Thalmann}, {Inhester},
  {Tadesse}, {Sun}, \& {Hoeksema}}]{2012SoPh..281...37W}
{Wiegelmann}, T., {Thalmann}, J.~K., {Inhester}, B., {et~al.} 2012, \solphys,
  281, 37

\bibitem[{{Zhang} {et~al.}(2022){Zhang}, {Musset}, {Glesener}, {Panesar}, \&
  {Fleishman}}]{Zhang2022}
{Zhang}, Y., {Musset}, S., {Glesener}, L., {Panesar}, N., \& {Fleishman}, G.
  2022, arXiv e-prints, arXiv:2207.05668

\end{thebibliography}

\begin{appendix}

\begin{figure*}[h!]
\centering
\includegraphics[width=0.99\textwidth]{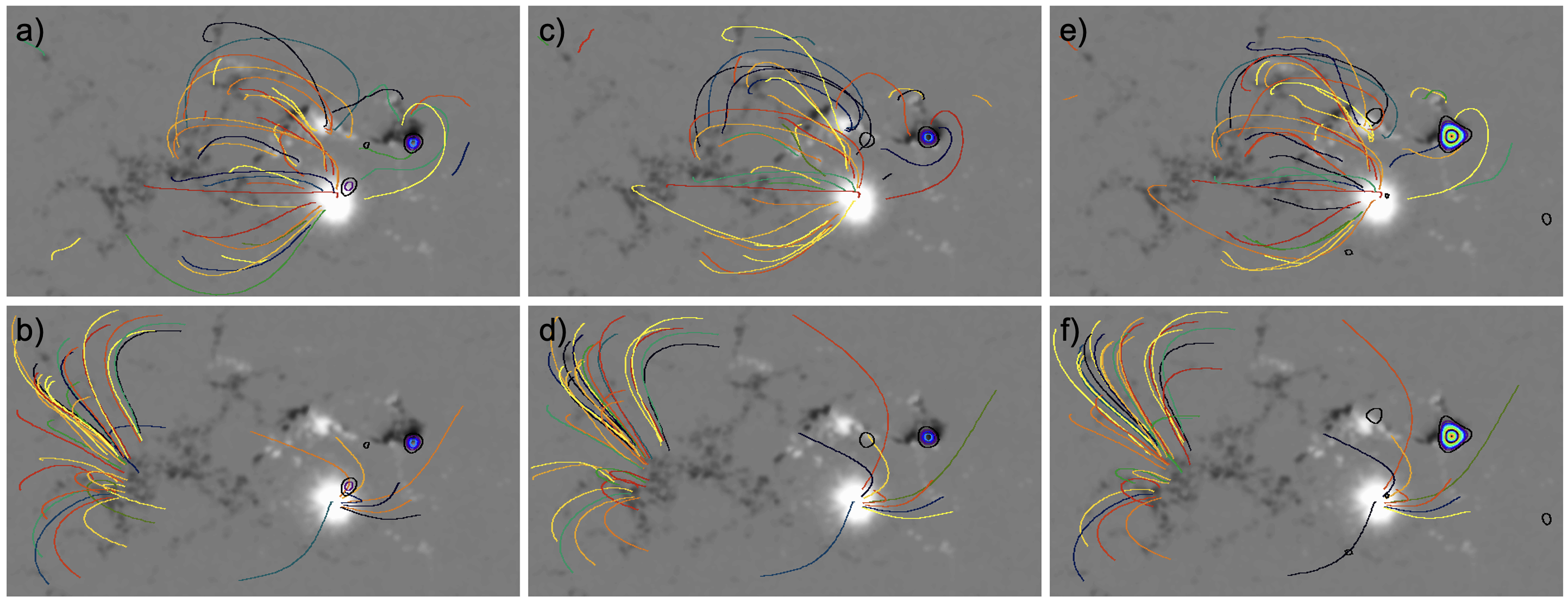}
\caption{Nonlinear force-free field extrapolations
based on SDO/HMI vector magnetograms and STIX overplotted
(see text for details).
\label{fig:nlfff} }
\end{figure*}

\section{NLFFF extrapolations of AR magnetic field topology}\label{appendix:NLFFF}


%
In order to understand the magnetic topology of eruptive event, we extrapolate photospheric vector magnetograms of the source AR obtained from SDO/HMI into the solar corona under the nonlinear force-free field (NLFFF) assumption with the help of an optimization code, as described in \cite{2012SoPh..281...37W}. The photospheric magnetic vector field measurements have been preprocessed to obtain consistent boundary data for the NLFFF-model (see \citealt{2006SoPh..233..215W} for details). 

Figure~\ref{fig:nlfff} depicts a sample of the extrapolated field lines, which are plotted over the vertical photospheric magnetic field shown in greyscale.  Additionally, we overplot the nonthermal STIX sources as coloured contours. Note that we have rotated the STIX images to conform with the vantage point of SDO, which is exactly the inverse process as used for Fig.~\ref{fig:stix_images} where we have rotated SDO/AIA images to the SolO view. We can do perform this rotation only for the nonthermal STIX sources since they originate in the chromosphere, while the thermal source is an extended 3D structure in the corona for which a simple rotation would cause severe projection effects.

We show this comparison for three of the nonthermal STIX peaks for which nearly co-temporal HMI data was available (times are UT at 1~AU): HMI at 06:24 and the STIX peak around 06:31~UT (panels a +b ), HMI at 06:36 and STIX around 06:37~UT (panels c +d ), and finally HMI at 06:48~UT and STIX at 06:46~UT (panels e + f). The top row of panels (a, c, e) shows closed loops where both footpoints of the magnetic field line are anchored in the photosphere. The bottom panels (b, d, f) contain open magnetic field lines, where open means that the field lines have only one footpoint in the photosphere and the upper end reaches the top boundary of the computational domain. These are the field lines which potentially offer access to interplanetary space.

\section{Directivity of type III bursts} \label{Sec:appendix_radio}

\begin{figure*}[h!]
\centering
\includegraphics[width= 0.99\textwidth]{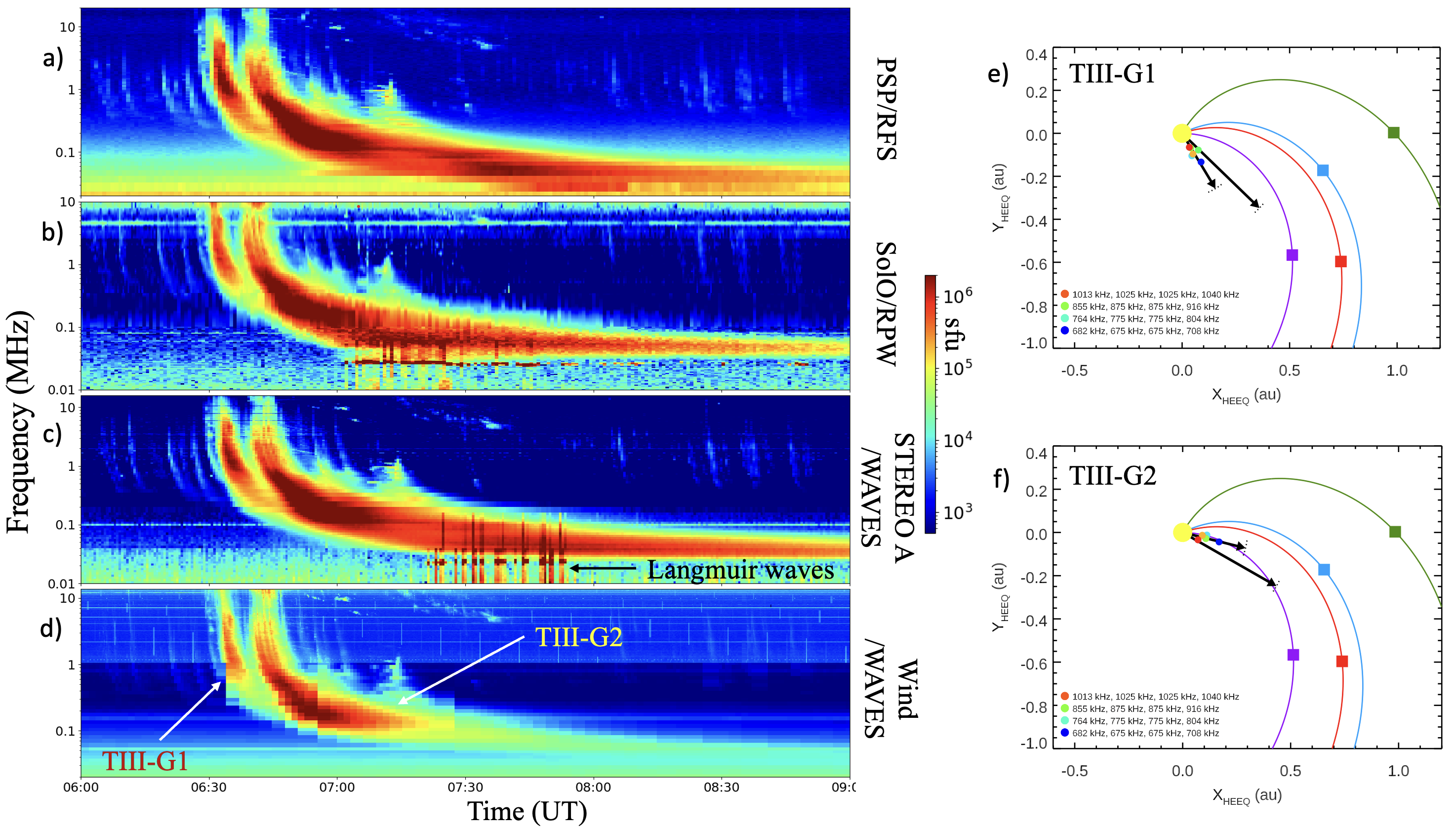}
\caption{Space-borne radio dynamic spectrograms from: (a) \textit{PSP}, (b) \textit{SolO}, (c) \textit{STEREO-A}, and (d) \textit{Wind}. Panels (e) and (f) show the direction finding analysis of TIII-G1 and TIII-G2.  \label{fig:space_obs_all} }
\end{figure*}

The interplanetary radio emissions associated with the event were observed by all spacecraft: \textit{STEREO-A}/WAVES, \textit{Wind}/WAVES, \textit{SolO}/RPW, and \textit{PSP}/FIELDS. A combined dynamic radio spectrum from the three spacecraft excluding \textit{WIND} is presented in Fig.~\ref{fig:space_obs_all}. The figure also includes the linear polarisation measurements from \textit{PSP} and \textit{STEREO-A}, which also provides goniopolarimetric measurements. 

Locating the source and the propagation of the type III radio sources in the corona and interplanetary space without interferometric imaging \citep[e.g., LOw Frequency ARray (LOFAR);][]{Haarlem13} or radio triangulation \citep[e.g.,][]{Magdalenic14, Krupar16, Jebaraj20} is challenging. However, we make use of the direction finding observations from \textit{STEREO-A} and \textit{Wind} observations \citep[azimuth and co-latitude angles of the radio emission, ][]{Krupar14a}. 


In the case of the directivity analysis, we have compared calibrated radio fluxes measured by Parker Solar Probe, Solar Orbiter, STEREO~A and \textit{Wind} at six frequency channels. We assumed that the radio emission pattern $S$ as a function of heliocentric longitude $\lambda$ can be described by the von Mises distribution (also known as the circular normal distribution) as:
\begin{equation}
S(\lambda)=\frac{\exp(\kappa\cos(\lambda-\lambda_0))}{2\pi I_0(\kappa)}
\mbox{,}
\label{eq:lambda}
\end{equation}

where $\lambda_0$ is a direction corresponding to a peak radio flux, $\kappa$ is a measure of concentration, and $I_0$ is the modified Bessel function of the first kind of order 0, with this scaling constant chosen so that the distribution sums to unity. 


Fig~\ref{fig:space_obs_all} panels e \& f show the propagation direction of TIII-G1 and TIII-G2 sources obtained using the radio triangulation technique \citep[short black arrows, for more details see ][]{Krupar14a} and a directivity analysis (long black arrows, radio emission pattern is described by the von Mises distribution). Both methods employ cross-calibrated measurements from multiple spacecraft, all four spacecraft in the case of the directivity analysis, \textit{Wind} and \textit{STEREO-A} in the case of radio triangulation. We perform both analyses to better constrain the direction of source propagation and eliminate intrinsic errors which may arise from each method. Even so, we shall put more weight on the radio triangulation results as they do not depend on free parameters such as a coronal electron density model.



The radio triangulation results of TIII-G1 between the frequencies 1000 kHz and 600 kHz indicate that the source propagated in the -58$^{\circ}$ heliolongitude (possible error of $\pm$6$^{\circ}$). The directivity analysis indicated that the source propagated in -43$^{\circ}$ heliolongitude (possible error of $\pm$2$^{\circ}$). Both results indicate that TIII-G1 most likely propagated westward of \textit{STEREO-A} close to the reference line (black dashed line) plotted in Fig.~\ref{fig:Multy_sc_Electron}. The closest spacecraft to the propagation of TIII-G1 was \textit{PSP}, however, no in-situ Langmuir waves were recorded by \textit{PSP}. Therefore, we can conclude that it was highly unlikely for TIII-G1 to be associated with the in-situ electrons recorded by \textit{SolO}, and \textit{STEREO-A}. \cite{Lario22} found a very impulsive rise of particle flux at \textit{PSP} during the early stages of the event, which is to be expected when there was direct connectivity in the direction rather close to \textit{PSP}.

For TIII-G2, we performed a similar analysis (Fig.~\ref{fig:space_obs_all}e) and found, according to radio triangulation, the source propagated in the -15$^{\circ}$ heliolongitude with a possible error of $\pm$2$^{\circ}$. Using the directivity analysis, we found the propagation to be in the heliolongitudes -30$^{\circ}$ with possibly a $\pm$3$^{\circ}$ error. The larger spread in propagation direction between radio triangulation and the directivity analysis could be due to errors arising from the propagation of radio waves in the presence of large-scale density fluctuations \citep[][]{Krupar20}. Alternatively, the large spread may also likely be due to multiple electron beams generating type III radio bursts within TIII-G2. The electron beams that generated the different type III bursts may propagate in slightly different solar wind due to their time-varying injections at the source \-- causing a large spread in their directivity. However, it should be noted that TIII-G2 propagated mostly in the direction of \textit{SolO} and \textit{STEREO-A}, which is further corroborated by the in-situ Langmuir waves. Therefore, from our analysis, TIII-G2 may be associated with the energetic electrons recorded in-situ at both \textit{SolO} and \textit{STEREO-A}.

\section{Shock wave kinematics} \label{Sec:appendix_shock}

We reconstructed the 3D structure of the shock wave using \textit{PyThea}, a software package to reconstruct the 3D structure of CMEs and shock waves \citep{Kouloumvakos2022}, written in Python language and available online\footnote{\url{https://doi.org/10.5281/zenodo.5713659}\label{Pythea}}. We took advantage of the two viewpoints provided by \textit{STEREO-A} and near-Earth spacecraft (i.e., \textit{SOHO} and \textit{SDO}), and we fitted an ellipsoid model to EUV and WL observations of the shock wave. We adjusted the free parameters of the geometrical ellipsoid model to achieve the best visual fit to near-simultaneous observations for the two available viewpoints. From the 3D reconstruction, we determined the position and kinematics of the shock in the corona.

\begin{figure*}
\centering
\includegraphics[width=0.85\textwidth]{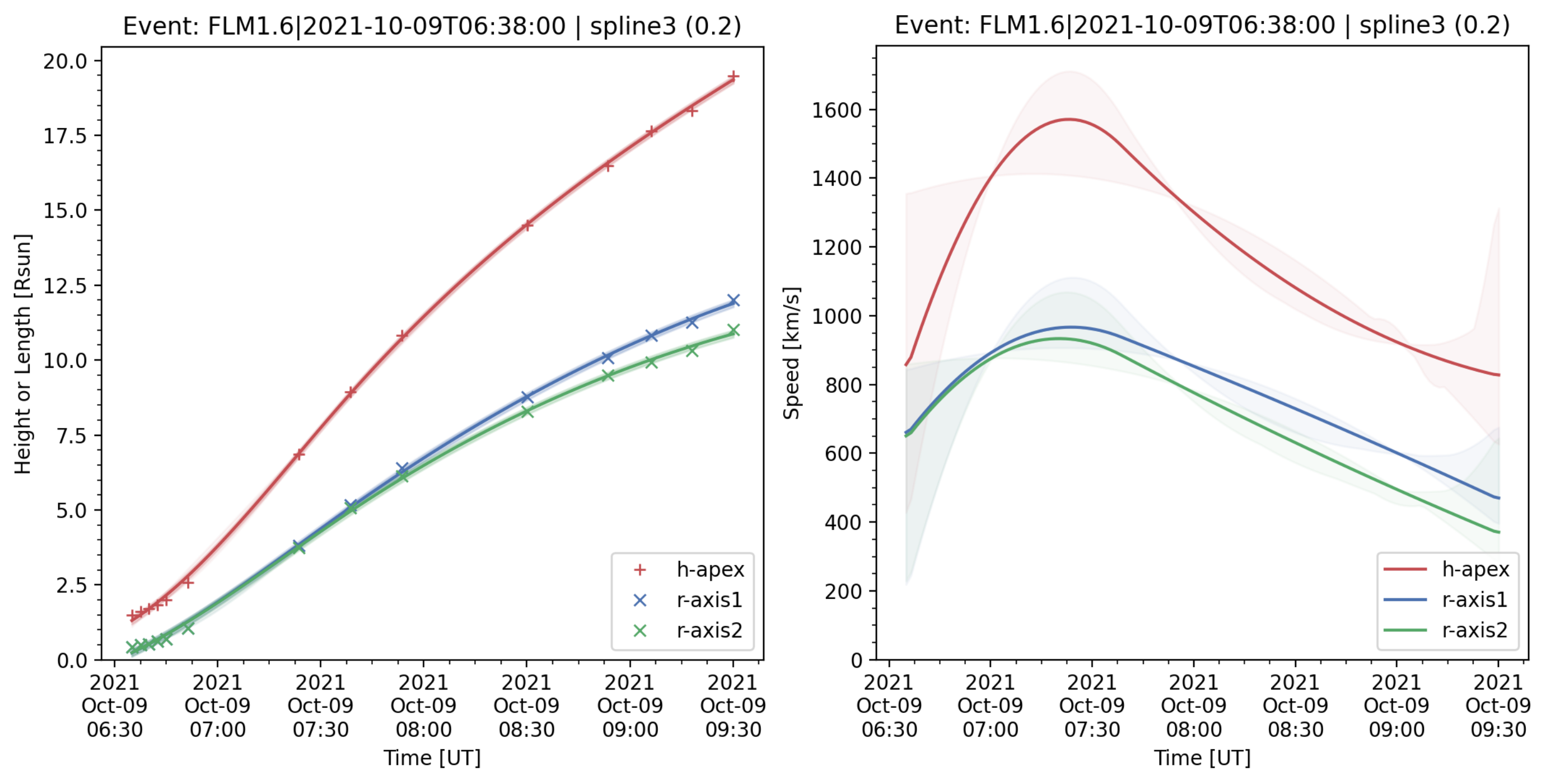}
\caption{ The kinematics of the reconstructed shock wave. The three axes of an ellipsoidal fit, namely, the shock apex (red), and the two flank axes (r-axis 1 - blue and r-axis 2 - green) are plotted together and represented with different colours. The left and the right panel of the figure show the height-time evolution of all the three shock axes, and its first derivative (velocity), respectively. The possible errors are in each axis and indicated by the shaded regions. The "+" and "x" markings on the curves are the constraints provided by the observations. \label{fig:shock_kinematics} }
\end{figure*}

Fig.~\ref{fig:shock_kinematics} suggests that the wave propagated quite rapidly at the apex and reached a maximum possible speed of 1400~km~s$^{-1}$ at around 07:25~UT. The shock apex was at a radial distance of 7~R$_{\odot}$ at this time. The flank regions of the shock expanded slower and reached their peak speeds of 900~km~s$^{-1}$ a little earlier at 07:15~UT when they were at a radial distance of 3~R$_{\odot}$. According to the ellipsoidal fit, both flanks expanded at a similar rate till their respective peaks after which their velocities were slightly different.   

According to the analysis of the EUV wave and the WL shock wave, the apex of the shock expanded considerably faster than the flank regions. Comparing these with the kinematics of the type II radio burst, we may get a first-order understanding of their association. To do so, we converted the spectral drift rate of the type II radio bursts into km~s$^{-1}$ using radial coronal electron density models. We applied a 5-fold \cite{Newkirk61} model to the metric and decametric type II bursts, and a 3.5-fold \cite{Saito70} model for the hectometer bursts. These coronal electron density models were chosen based on the projection analysis of type II radio sources at shock flanks and for non-radial shock propagation done by \cite{Jebaraj20}. We obtained 1100$\pm$30~km~s$^{-1}$ for TII(1a) and TII(1b), which suggests that the emission may have been from regions between the flanks and the apex. Meanwhile, for TII(2a) and TII(2b), using a similar approach, we obtained a drift speed of 700$\pm$50~km~s$^{-1}$ associating it with the flank regions of the shock wave. The $\pm$ values here have been calculated from the varying bandwidth of the respective type II bursts over time. TII(2b) on the other hand was treated separately due to its prolonged emission period in the hectometer wavelength. We obtained a drift speed of 750$\pm$50~km~s$^{-1}$, which corresponds well to the flank regions of the shock wave during this time. Meanwhile, TII(HB) and TII(W) are not as straightforward to analyse using a simple radial density profile such as the one applied here. The two structured type II bursts were most likely due to shock wave propagation in regions of enhanced density and therefore exhibit spectral deformities.

\end{appendix}

\end{document}